\documentclass[11pt]{article}
\usepackage[utf8]{inputenc}
\usepackage[T1]{fontenc}

\usepackage[english,activeacute]{babel}
\usepackage{natbib}
\usepackage{comment}
\usepackage{float}
\usepackage[hidelinks]{hyperref}
\usepackage{mathrsfs}
\usepackage{enumitem}
\usepackage[font={small,it}]{caption}
\usepackage{amsmath,amsfonts,amsthm,amssymb}
\usepackage{bm,rotating,multirow,dsfont,graphicx}
\usepackage[usenames, dvipsnames]{color}
\usepackage{url}
\usepackage{multicol}
\usepackage{multirow}
\usepackage[T1]{fontenc}
\usepackage{flafter}
\usepackage{appendix}
\usepackage{subfigure}
\usepackage{xcolor}
\usepackage{soul}
\usepackage{setspace}
\usepackage{booktabs}
\usepackage{tabularx}
\usepackage{setspace}
\usepackage{subcaption}

\usepackage{algorithm}
\usepackage{algpseudocode}

\usepackage{siunitx}

\usepackage{tikz}
\usetikzlibrary{arrows.meta}
\usetikzlibrary{positioning}
\usetikzlibrary{calc}

\makeatletter
\def\hlinewd#1{%
	\noalign{\ifnum0=`}\fi\hrule \@height #1 %
	\futurelet\reserved@a\@xhline}
\makeatother
\def\spacingset#1{\renewcommand{\baselinestretch}{#1}\small\normalsize}
\spacingset{1}

\newcommand\simind{\mathrel{\overset{\makebox[0pt]{\mbox{\normalfont\tiny\sffamily ind}}}{\sim}}}

\begin{document}

\title{\textbf{Uncovering latent territorial structure in \vspace{0.2cm} \\ ICFES Saber 11 performance with \vspace{0.2cm} \\ Bayesian multilevel spatial models}}

\date{}

\author{
    Laura Pardo\qquad Juan Sosa\footnote{Corresponding author: jcsosam@unal.edu.co.}\\ Juan Pablo Torres-Clavijo\qquad Andrés Felipe Arévalo-Arévalo\\
    \vspace{0.1cm} \\
    Universidad Nacional de Colombia, Colombia 
}

\maketitle

\begin{abstract}
This article develops a Bayesian hierarchical framework to analyze academic performance in the 2022 second semester Saber 11 examination in Colombia. Our approach combines multilevel regression with municipal and departmental spatial random effects, and it incorporates Ridge and Lasso regularization priors to compare the contribution of sociodemographic covariates. Inference is implemented in a fully open source workflow using Markov chain Monte Carlo methods, and model behavior is assessed through synthetic data that mirror key features of the observed data. Simulation results indicate that Ridge provides the most balanced performance in parameter recovery, predictive accuracy, and sampling efficiency, while Lasso shows weaker fit and posterior stability, with gains in predictive accuracy under stronger multicollinearity. In the application, posterior rankings show a strong centralization of performance, with higher scores in central departments and lower scores in peripheral territories, and the strongest correlates of scores are student level living conditions, maternal education, access to educational resources, gender, and ethnic background, while spatial random effects capture residual regional disparities. A hybrid Bayesian segmentation based on K means propagates posterior uncertainty into clustering at departmental, municipal, and spatial scales, revealing multiscale territorial patterns consistent with structural inequalities and informing territorial targeting in education policy.
\end{abstract}


\spacingset{1.1} 

\newpage

\section{Introduction}

The Saber 11 test is a standardized external assessment administered by the Colombian Institute for the Evaluation of Education (ICFES, by its acronym in Spanish) and it is part of the Colombian national evaluation system. The exam monitors the development of competencies of students who are about to complete eleventh grade, it provides information on the quality of education offered by institutions across the country, and it informs public education policies at the national, territorial, and institutional levels \cite{decreto869}. This paper studies the global score, defined as a weighted average of the five assessed components, Mathematics, Reading, Science, Social Studies, and English.

Academic performance is shaped by contextual conditions that operate beyond the classroom. Access to safe drinking water, adequate sanitation, exposure to violence, and basic living conditions can determine whether sustained learning is feasible. Colombia has experienced a prolonged internal conflict and persistent territorial inequalities, which have contributed to unequal state capacity and uneven delivery of public services across regions. Quantifying how these structural conditions relate to educational outcomes helps identify where policy and resources should be redirected, and it clarifies how territorial disparities translate into gaps in learning opportunities.

Recent evidence indicates that educational outcomes in Colombia exhibit spatial dependence and regional clustering, which suggests that place based mechanisms act as structural determinants of educational quality \citep{rodriguez2022spatial}. This motivates modeling strategies that move beyond purely individual level covariates and that incorporate latent territorial structure. Bayesian hierarchical models are well suited for this goal because they borrow strength across groups, mitigate overfitting, and provide principled uncertainty quantification in complex multilevel settings \citep{Gelman2014,Sosa2022BayesianHierarchical}. Prior work has also supported Bayesian approaches for educational outcomes, including models designed to bounded responses \citep{cepeda2016spatial}. Recent developments in spatially varying effects and structured shrinkage further strengthen the case for flexible spatial inference in settings where spatial structure and multicollinearity can coexist \citep{liu2024generalized,sakai2024bayesian}.

This study develops a Bayesian hierarchical framework to analyze determinants of academic performance in the 2022 second semester Saber 11 examination. The framework combines multilevel regression with departmental and municipal components, and it includes a spatial random effect that captures latent territorial structure not explained by observed covariates. The model links student scores to a design matrix and regression coefficients through a linear predictor that includes hierarchical and spatial terms, and it targets fundamental posterior summaries for uncertainty aware inference and decision support. The spatial component encodes geographic proximity and residual dependence, following standard Bayesian spatial modeling principles as in \cite{Besag1991BYM}, \cite{AG25p}, and \cite{Reich2019Bayesian}.

A central methodological contribution is the integration of regularization through Ridge and Lasso priors (e.g., \citealt{Tibarti} and \citealt{Gillariose2025LassoRidge}), which supports estimation under correlated predictors and enables a principled contrast of the relative contribution of sociodemographic covariates. The Ridge specification favors global shrinkage, while the Lasso specification induces stronger sparsity, and this comparison clarifies when shrinkage improves posterior stability, interpretability, and predictive performance. Model behavior is evaluated using synthetic data designed to reflect key features of the observed data, including increasing dependence among predictors, and the fitted models are then applied to the 2022 second semester Saber 11 data to quantify covariate effects and latent territorial structure.

In the case study, the paper addresses four important questions. First, which student level living conditions and institutional characteristics are most strongly associated with the global score after accounting for hierarchical structure. Second, to what extent territorial covariates explain performance gaps once student level conditions are included. Third, how much residual territorial structure remains after conditioning on covariates, and how this structure is captured by the spatial random effects. Fourth, how the choice between Ridge and Lasso regularization affects parameter recovery, posterior stability, and predictive performance across controlled simulations and real data.

The empirical results show a marked centralization of educational performance. Posterior rankings place the highest scores in central departments, including Bogot\'a, Boyac\'a, and Cundinamarca, and the lowest scores in peripheral territories, including Choc\'o, Vaup\'es, and Vichada. Student level living conditions exhibit the strongest associations with performance, particularly mother education and access to educational resources, and the results also show systematic gaps associated with gender and ethnic background. Territorial covariates have smaller marginal effects, yet spatial random effects capture persistent regional disparities not explained by observed covariates, which supports the conclusion that unequal opportunities remain even after adjusting for measured conditions.

A second methodological contribution is a hybrid Bayesian segmentation strategy that propagates posterior uncertainty into clustering. We develop a $K$ means based procedure to produce multiscale territorial clusters at the departmental and municipal levels, and from the spatial random effects, which yields a latent segmentation aligned with structural inequalities and reveals heterogeneity within low performing departments. In practical terms, this segmentation complements posterior rankings by identifying territorially coherent groups that may benefit from differentiated policy targeting and resource allocation. Finally, this work provides a fully open source theoretical and computational implementation, with custom Markov chain Monte Carlo algorithms that embed Metropolis--Hastings updates within a Gibbs sampler, and with model comparison based on posterior predictive checks and information criteria \citep{Gelman2014}. This end to end workflow is designed to be adapted to future educational cohorts and to related applications where outcomes are shaped by multilevel organization and latent spatial structure.

The remainder of the document is organized as follows. Section 2 describes the Saber 11 dataset, the student, municipal, and departmental covariates, and their spatial distribution. Section 3 presents the Bayesian multilevel spatial regression specifications, prior choices, and posterior computation. Section 4 reports posterior inference, territorial rankings and segmentation, model comparison, and out of sample prediction. Section 5 presents the simulation design and results. Section 6 summarizes the main findings and outlines directions for future research.

\begin{figure}[!htb]
    \centering
    \subfigure[Departmental level mean.]
    {\includegraphics[width=0.49\textwidth,trim={2cm 0cm 2cm 0cm},clip]{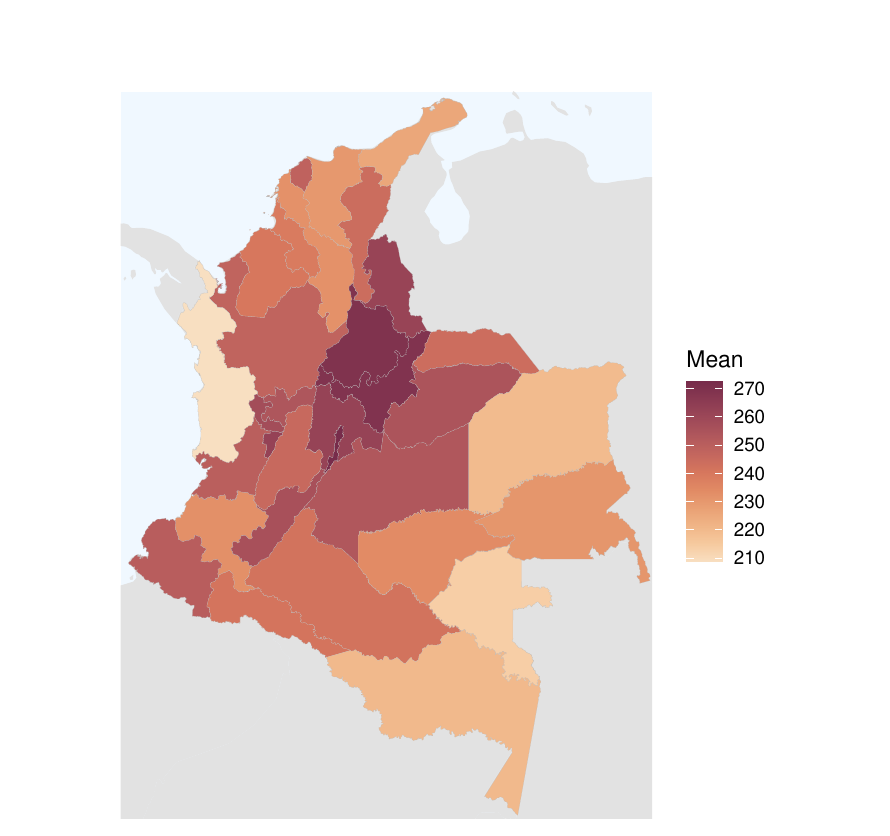}}
    \subfigure[Municipal level mean.]
    {\includegraphics[width=0.49\textwidth,trim={2cm 0cm 2cm 0cm},clip]{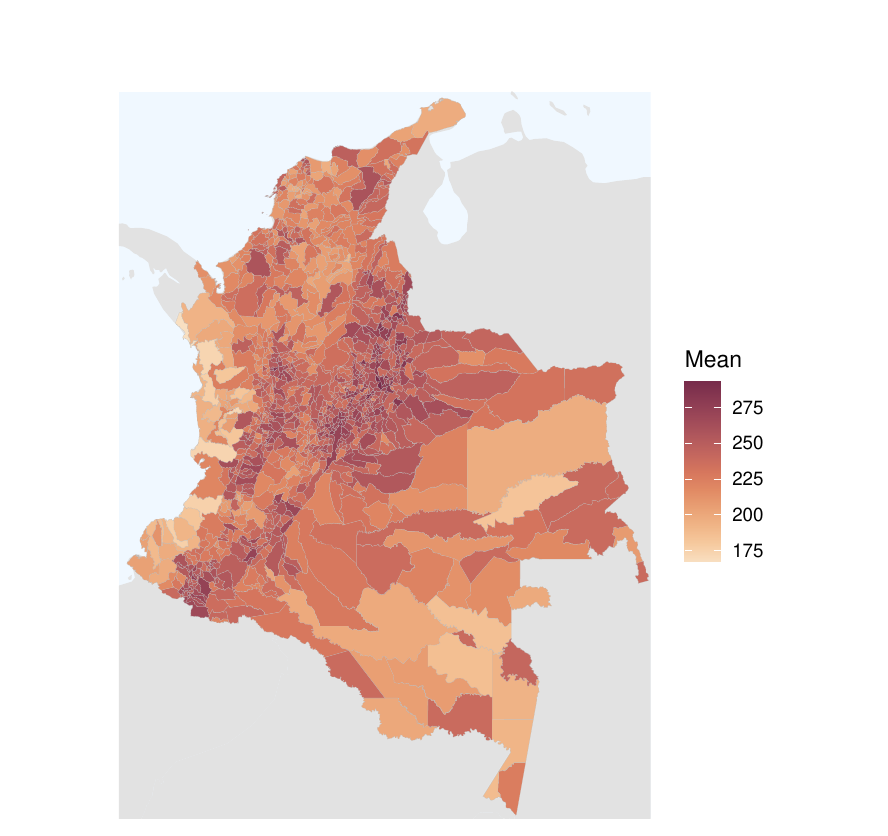}}
    \caption{Average Saber 11 global scores by municipality and department.}
    \label{fig:comparacion_global_score}
\end{figure}

\section{Saber 11 data}\label{sec_eda}

In this article we propose three Bayesian hierarchical regression models to study the Saber 11 global score in the second semester of 2022. The models incorporate sociodemographic covariates to assess how student characteristics and contextual factors relate to overall performance. In this section we provide a descriptive overview of the global score across the Colombian territory, we list the covariates included in the analysis, and we present the spatial distribution of selected covariates.

To contextualize the analysis, this section examines descriptive statistics of the global score and the sociodemographic variables used to develop the global score models. The analysis considers 22 covariates describing departmental, municipal, and student contexts, including numerical and categorical variables. As shown in panel (a) of Figure \ref{fig:comparacion_global_score}, higher scores concentrate in the central region of the country, and average performance decreases as departments are located farther from the center. This pattern is also present at the municipal level, as shown in panel (b) of Figure \ref{fig:comparacion_global_score}, where some municipalities exhibit mean global scores that exceed the average score of their corresponding departments.

\subsection{Departmental level}

Table \ref{tab_eda_dep} reveals marked variability across the departmental covariates. GDP exhibits a wide range, with a mean of 28.52 million COP and a high coefficient of variation of 51.7\%. The rural population rate also shows substantial dispersion, ranging from nearly 0\% to fully rural departments. The percentage of municipalities at risk has the highest variability, with a coefficient of variation of 70.7\%, which reflects strong heterogeneity in risk exposure. The homicide rate also shows considerable dispersion, with a coefficient of variation of 52.3\%, which indicates meaningful differences in violence levels across departments.

\begin{table}[!htb]
\centering
\resizebox{\textwidth}{!}{%
\begin{tabular}{m{4cm} m{5.8cm} c c c c c c c}
\toprule
Variable & Description & Min & Max & $Q_{1}$ & $Q_{2}$ & $Q_{3}$ & Mean & CV (\%) \\
\midrule
Department GDP per capita & Gross domestic product per capita (million COP) & 8.00 & 50.27 & 22.49 & 27.45 & 31.20 & 28.52 & 51.7 \\
Rural population share & Percentage of the population living in rural areas & 0.05 & 100 & 1.66 & 11.84 & 39.70 & 23.02 & 47.9 \\
Municipalities at risk & Percentage of municipalities with some level of risk & 0.00 & 100 & 18.92 & 47.62 & 64.00 & 46.37 & 70.7 \\
Weighted homicide rate & Homicide rate per 100{,}000 inhabitants, population weighted & 6.89 & 53.67 & 15.25 & 18.68 & 23.26 & 21.79 & 52.3 \\
\bottomrule
\end{tabular}%
}
\caption{Descriptive summary of department level variables.}
\label{tab_eda_dep}
\end{table}

\subsection{Municipal level}

Table \ref{tab_eda_mun} exhibits substantial heterogeneity across municipalities. Indicators such as terrorism, theft, homicides, and kidnapping have very high coefficients of variation, which reflects strong dispersion and the presence of municipalities where violence is highly concentrated relative to most others. In contrast, the percentage of public school students shows low variability. The teacher to student ratio, victimization risk, and distance to the departmental capital exhibit moderate to high dispersion across municipalities. Furthermore, the departmental maps, not shown here, for GDP (million COP), rural population (\%), municipalities at risk (\%), and homicides per 100{,}000 inhabitants, together with the municipal maps, not shown here, for the victimization risk index and distance to the departmental capital (km), that summarize the spatial distribution of selected departmental and municipal covariates across the country, reveal clear patterns of resource concentration in central regions and a markedly uneven distribution of violence across the territory.

\begin{table}[!htb]
\centering
\resizebox{\textwidth}{!}{%
\begin{tabular}{m{3.8cm} m{5.8cm} c c c c c c c}
\toprule
Variable & Description & Min & Max & $Q_{1}$ & $Q_{2}$ & $Q_{3}$ & Mean & CV (\%) \\
\midrule
Teacher to student ratio & Ratio of teachers to students & 0.03 & 0.79 & 0.05 & 0.05 & 0.06 & 0.05 & 47.6 \\
Victimization risk & Victimization risk index (2022) & 0.02 & 1.00 & 0.12 & 0.14 & 0.20 & 0.19 & 84.2 \\
Homicides & Homicide rate per 100{,}000 inhabitants & 0.00 & 302.51 & 13.61 & 17.86 & 26.15 & 24.40 & 128.4 \\
Public school students & Percentage of students enrolled in public schools & 23.18 & 100.00 & 64.50 & 77.93 & 92.20 & 78.44 & 10.2 \\
Terrorism & Reported cases of terrorism & 0.00 & 29.00 & 0.00 & 0.00 & 1.00 & 0.60 & 686.5 \\
Theft & Reported cases of theft & 0.00 & 172{,}082 & 163 & 2{,}146 & 15{,}931 & 31{,}369.5 & 636.4 \\
Kidnapping & Number of registered kidnappings & 0.00 & 6.00 & 0.00 & 0.00 & 2.00 & 1.07 & 411.7 \\
Distance to the capital & Distance to the departmental capital (approx.\ km) & 0.00 & 376.12 & 0.00 & 13.97 & 63.62 & 40.55 & 72.3 \\
\bottomrule
\end{tabular}%
}
\caption{Descriptive summary of municipal level variables.}
\label{tab_eda_mun}
\end{table}

\subsection{Student level}

For each student, 10 categorical variables are included, which correspond to 21 columns in the design matrix. The proportion of each category is reported in Table \ref{tab_eda_stu}. The student level variables show clear patterns in access to resources and socioeconomic conditions. Most students report internet and computer access at home, while a smaller share report that their mother has higher education or that they belong to an ethnic group. Socioeconomic levels concentrate in the low and middle categories, with very few students in the highest levels. Nearly all students are enrolled in Calendar A schools, and 23\% attend private institutions. Regarding the study work balance, most students do not work or work fewer than 10 hours per week, and only a small share work more than 20 hours.

\begin{table}[!htb]
\centering
\resizebox{0.9\textwidth}{!}{%
\begin{tabular}{m{4cm} m{2.5cm} m{6.5cm} l}
\toprule
Variable & Levels & Level description & Percentage \\
\midrule
Mother education & Yes & Mother education is at least undergraduate & 14.8\% \\
\midrule
Computer & Yes & Has a computer at home & 54.62\% \\
\midrule
Internet & Yes & Has internet access at home & 73.2\% \\
\midrule
Ethnicity & Yes & Belongs to an ethnic group & 0.66\% \\
\midrule
Gender & Female & Female & 54.24\% \\
\midrule
Number of books at home & books\_11\_25 & Between 11 and 25 books at home & 30.71\% \\
& books\_26\_100 & Between 26 and 100 books at home & 17.85\% \\
& books\_more100 & More than 100 books at home & 0.44\% \\
\midrule
Socioeconomic level & 1 & Socioeconomic level 1 & 29.57\% \\
& 2 & Socioeconomic level 2 & 36.40\% \\
& 3 & Socioeconomic level 3 & 22.2\% \\
& 4 & Socioeconomic level 4 & 5.62\% \\
& 5 & Socioeconomic level 5 & 1.7\% \\
& 6 & Socioeconomic level 6 & 0.75\% \\
\midrule
School calendar & Calendar A & Takes the test in the second semester & 99.6\% \\
& Calendar B & Takes the test in the first semester & 0.13\% \\
\midrule
School type & Private & Private school & 23\% \\
\midrule
Student work (weekly hours) & Less than 10 & Works less than 10 hours per week & 22\% \\
& 11 to 20 & Works between 11 and 20 hours per week & 9.17\% \\
& 21 to 30 & Works between 21 and 30 hours per week & 3.17\% \\
& More than 30 & Works more than 30 hours per week & 3.33\% \\
\bottomrule
\end{tabular}%
}
\caption{Descriptive summary of student level covariates.}
\label{tab_eda_stu}
\end{table}

\section{Modeling}\label{sec_modeling}

The proposed models in this section are based on a fully Bayesian linear regression framework that incorporates a municipality level spatial random effect to capture territorial variation. Spatial dependence is modeled using a CAR distribution \citep{AG25p}. Several model variants are obtained by placing different shrinkage priors on the regression coefficients, with the aim of comparing their performance.

\subsection{Regularization}\label{sec_regu}

Consider the regression setting $\mathbf{y}=\mathbf{X}\boldsymbol{\beta}+\boldsymbol{\epsilon}$, where $\boldsymbol{\beta}\in\mathbb{R}^{p+1}$ is the vector of unknown regression coefficients. The ordinary least squares (OLS) estimator $\hat{\boldsymbol{\beta}}_{\mathrm{OLS}}$ minimizes the sum of squared residuals $\textsf{SSR}(\boldsymbol{\beta})=\lVert \mathbf{y}-\mathbf{X}\boldsymbol{\beta}\rVert_2^2$. Shrinkage methods modify this criterion by imposing constraints or penalties on $\boldsymbol{\beta}$ to regularize estimation, that is, to reduce variance and improve stability and prediction, particularly under multicollinearity. See \cite{Tibarti} and \cite{Gillariose2025LassoRidge} for an in depth discussion of regularization in the classical setting.

\subsubsection{Ridge shrinkage}

Ridge regularization estimates $\boldsymbol{\beta}$ by minimizing the sum of squared residuals subject to an $\ell_2$ constraint, equivalently by minimizing the penalized criterion
\[
\ell_{\mathrm{Ridge}}(\boldsymbol{\beta})
= \lVert\mathbf{y}-\mathbf{X}\boldsymbol{\beta}\rVert_2^2
+ \lambda\lVert\boldsymbol{\beta}\rVert_2^2,
\]
where $\lVert\boldsymbol{\beta}\rVert_2^2=\sum_{j=1}^{p}\beta_j^2$ and $\lambda>0$ controls the amount of shrinkage. This penalty shrinks coefficients continuously toward zero without setting them exactly to zero, which improves stability under multicollinearity. In practice, the intercept is typically excluded from penalization, and $\lambda$ is commonly selected by cross validation.

In the Bayesian formulation, Ridge shrinkage arises by assigning a zero centered Gaussian prior to the regression coefficients, $\beta_j \mid \lambda^2 \overset{\text{ind}}{\sim} \textsf{N}(0,\lambda^{-2})$ for $j=1,\ldots,p$, with $\lambda^2 \sim \textsf{G}(a_\lambda,b_\lambda)$ and $\sigma^2 \sim \textsf{IG}(a_\sigma,b_\sigma)$ (see Appendix \ref{app_proofs}). Under a Gaussian likelihood, the posterior mode of $\boldsymbol{\beta}$ coincides with the classical Ridge estimator, and the hierarchical prior treats $\lambda^2$ as an unknown global precision that adapts the amount of shrinkage to the data while providing posterior uncertainty for coefficients and predictions \citep{PerezElizalde2022HDBRR}.

\subsubsection{Lasso shrinkage}

Lasso regularization estimates $\boldsymbol{\beta}$ by minimizing the sum of squared residuals under an $\ell_1$ constraint, equivalently by minimizing the penalized criterion
\[
\ell_{\mathrm{Lasso}}(\boldsymbol{\beta})
= \lVert\mathbf{y}-\mathbf{X}\boldsymbol{\beta}\rVert_2^2
+ \lambda\lVert\boldsymbol{\beta}\rVert_1,
\]
where $\lVert\boldsymbol{\beta}\rVert_1=\sum_{j=1}^{p}\lvert\beta_j\rvert$ and $\lambda>0$ controls the amount of shrinkage. Unlike Ridge, the $\ell_1$ penalty can set coefficients exactly to zero, which yields sparse solutions and supports variable selection, particularly when the design exhibits strong collinearity or when many predictors have weak effects.

In the Bayesian formulation, Lasso shrinkage arises by assigning independent Laplace priors to the regression coefficients, $\beta_j \mid \lambda \overset{\text{ind}}{\sim} \textsf{Laplace}(0,\lambda^{-1})$ for $j=1,\ldots,p$, with $\lambda \sim \textsf{G}(a_\lambda,b_\lambda)$ and $\sigma^2 \sim \textsf{IG}(a_\sigma,b_\sigma)$ (see Appendix \ref{app_proofs}). Under a Gaussian likelihood, the posterior mode of $\boldsymbol{\beta}$ coincides with the classical Lasso estimator, and posterior inference provides uncertainty quantification for both sparsity patterns and predictions. Computationally, the Laplace prior admits a convenient scale mixture representation, 
$$
\beta_j \mid \tau_j^2 \overset{\text{ind}}{\sim} \textsf{N}(0,\tau_j^2),\quad \tau_j^2 \mid \lambda^2 \overset{\text{ind}}{\sim} \textsf{Exp}\!\left(\lambda^2/2\right),
$$ 
which facilitates conditionally Gaussian updates in Gibbs type samplers (see Appendix \ref{app_proofs}).

\subsection{Conditional autoregressive model}\label{sec_car}

Let us review basic concepts for conditional autoregressive models (CAR; e.g., \citealt{AG25p}) before specifying the proposed models. Areal data models specify a joint distribution for a finite collection of random variables indexed by a partition of a region $S$ into areal units $s_1,\ldots,s_n$, where $y_i$ denotes the measurement for unit $s_i$. In this work, spatial dependence is incorporated as a random effect within a hierarchical mixed model. Spatial association is encoded through a neighborhood relation $\mathcal{R}$, where $s_i \sim_{\mathcal{R}} s_j$ when units share a boundary or vertex and $i \neq j$, which induces a symmetric binary adjacency matrix $\mathbf{W}$ with entries $W_{i,j}=1$ if $s_i \sim_{\mathcal{R}} s_j$ and $W_{i,j}=0$ otherwise. The diagonal matrix $\mathbf{D}$ summarizes local connectivity, with $D_{i,i}=d_i=\sum_{j=1}^{n} W_{i,j}$ and $D_{i,j}=0$ for $i \neq j$, and these matrices define the CAR prior for spatial random effects.

Building on this neighborhood structure, the intrinsic Gaussian CAR prior specifies each spatial effect $y_i$ conditionally on the effects in neighboring areal units, which induces spatial autocorrelation by shrinking $y_i$ toward the local average,
\[
y_i \mid \mathbf{y}_{\sim i}, \tau^2 \overset{\text{ind}}{\sim} \textsf{N}\!\left(\frac{1}{d_i}\sum_{j \sim i} y_j,\frac{\tau^2}{d_i}\right),
\]
where $\mathbf{y}_{\sim i}$ denotes the vector of  spatial effects of all neighboring areal units of $s_i$ excluding $y_i$, and $d_i=\sum_{j=1}^{n}W_{i,j}$. These full conditionals are compatible and, by Brook lemma \citep{AG25p}, they imply the joint distribution

\[
\mathbf{y}\mid \tau^2,\mathbf{W},\mathbf{D}\sim \textsf{N}_{n}\!\left(\mathbf{0},\tau^2(\mathbf{D}-\mathbf{W})^{-1}\right),
\]
where $\mathbf{y}=(y_1,\ldots,y_n)^{\top}$. This prior is intrinsic and improper because the precision matrix $\tau^{-2}(\mathbf{D}-\mathbf{W})$ is singular, since $(\mathbf{D}-\mathbf{W})\mathbf{1}=\mathbf{0}$, and therefore an identifiability constraint, such as a sum to zero constraint, is required to define a proper model (see Appendix \ref{app_proofs}).

\subsection{Normal model with mean and variance parameters and a spatial random effect}\label{sec_model1}

Let $y_{i,j,k}$ denote the Saber 11 global score of student $i$ in municipality $j$ of department $k$. We assume the sampling distribution
\begin{equation}\label{eq_vero_m1}
y_{i,j,k} \mid \zeta_{i,j,k}, \kappa_{j,k}^2 \overset{\text{ind}}{\sim} \textsf{N}\!\left(\zeta_{i,j,k}, \kappa_{j,k}^2\right),    
\end{equation}
for $i=1,\ldots,n_{j,k}$, $j=1,\ldots,m_k$, and $k=1,\ldots,d$. The mean includes a spatial random effect and it is defined as
\begin{align}\label{eq_mean_m1}
\zeta_{i,j,k}
&= \beta_0
+ \mathbf{x}_{i,j,k}^{\top}\boldsymbol{\beta}_{\mathrm{E}}
+ \mathbf{z}_{j,k}^{\top}\boldsymbol{\beta}_{\mathrm{M}}
+ \mathbf{w}_{k}^{\top}\boldsymbol{\beta}_{\mathrm{D}}
+ \phi_{j,k},
\end{align}
where $\mathbf{x}_{i,j,k}$, $\mathbf{z}_{j,k}$, and $\mathbf{w}_{k}$ are the covariate vectors at the student, municipal, and departmental levels, with dimensions $p_{\mathrm{E}}$, $p_{\mathrm{M}}$, and $p_{\mathrm{D}}$, respectively. Stacking the students in municipality $j$ of department $k$, the mean can be written in vector form as
\begin{align*}
\boldsymbol{\zeta}_{j,k}
&= \mathbf{1}_{n_{j,k}} \beta_0
+ \mathbf{X}_{j,k}\boldsymbol{\beta}_{\mathrm{E}}
+ \mathbf{1}_{n_{j,k}}\mathbf{z}_{j,k}^{\top}\boldsymbol{\beta}_{\mathrm{M}}
+ \mathbf{1}_{n_{j,k}}\mathbf{w}_{k}^{\top}\boldsymbol{\beta}_{\mathrm{D}}
+ \mathbf{1}_{n_{j,k}}\phi_{j,k},
\end{align*}
where $\mathbf{X}_{j,k} = [\mathbf{x}_{1,j,k}, \ldots, \mathbf{x}_{n_{j,k},j,k}]^{\top}$ denotes the matrix of student level covariates and $\mathbf{1}_{n_{j,k}}$ denotes a column vector of ones of length $n_{j,k}$.

\begin{figure}[!htb]
\centering
\resizebox{\textwidth}{!}{%

\begin{tikzpicture}[
    node distance=0.25cm,
    font=\huge, 
    every node/.style={font=\sffamily\large}, 
    hyper/.style={draw, rectangle, minimum size=0.8cm, align=center},
    param/.style={draw, circle, minimum size=0.9cm, align=center},
    obs/.style={draw, circle, fill=gray!30, minimum size=1.0cm, align=center},
    arrow/.style={-Latex} 
]


\node[hyper] (A) {$\nu_\beta$};
\node[hyper, right=of A] (B) {$\gamma_\beta^2$};
\node[hyper, right=of B] (C) {$\mu_\beta$};

\node[hyper, right=0.5cm of C] (D) {$\nu_\text{E}$};
\node[hyper, right=of D] (E) {$\gamma^2_\text{E}$};
\node[hyper, right=of E] (F) {$\mu_\text{E}$};

\node[hyper, right=0.5cm of F] (G) {$\nu_\text{M}$};
\node[hyper, right=of G] (H) {$\gamma^2_\text{M}$};
\node[hyper, right=of H] (I) {$\mu_\text{M}$};

\node[hyper, right=0.5cm of I] (X) {$\mathbf{X}$};

\node[hyper, right=0.5cm of X] (J) {$\nu_\text{D}$};
\node[hyper, right=of J] (K) {$\gamma^2_\text{D}$};
\node[hyper, right=of K] (L) {$\mu_\text{D}$};

\node[hyper, right=0.5cm of L] (M) {$W_k$};

\node[hyper, right=of M] (N) {$\nu_\phi$};
\node[hyper, right=of N] (O) {$\gamma^2_\phi$};

\node[hyper, right=0.5cm of O] (P) {$\nu_\kappa$};
\node[hyper, right=of P] (Q) {$a_{\alpha_\kappa}$};
\node[hyper, right=of Q] (R) {$b_{\alpha_\kappa}$};
\node[hyper, right=of R] (S) {$a_{\beta_\kappa}$};
\node[hyper, right=of S] (T) {$b_{\beta_\kappa}$};

\node[param, below=2cm of A, xshift=0.5cm, label=left:  \small IG] (SBETA) {$\sigma^2_\beta$};
\node[param, below=2cm of D, xshift=0.5cm, label=right:\small IG] (SE) {$\sigma^2_\text{E}$};
\node[param, below=2cm of G, xshift=0.5cm, label=right:\small IG] (SM) {$\sigma^2_\text{M}$};
\node[param, below=2cm of J, xshift=0.5cm, label=right:\small IG] (SD) {$\sigma^2_\text{D}$};
\node[param, below=2cm of N, xshift=0.5cm, label=right:\small IG] (TPHI) {$\tau^2_\phi$};
\node[param, below=2cm of Q, xshift=0.5cm, label=right:\small G] (AKAP) {$\alpha_\kappa$};
\node[param, below=2cm of S, xshift=0.5cm, label=right:\small G] (BKAP) {$\beta_\kappa$};

\node[param, below=1.5cm of SBETA, xshift=1cm, label=left:\small N] (BETA) {$\beta$};
\node[param, below=1.5cm of SE, xshift=1cm, label=left:\small N] (BE) {$\boldsymbol{\beta}_\text{E}$};
\node[param, below=1.5cm of SM, xshift=1cm, label=right:\small N] (BM) {$\boldsymbol{\beta}_\text{M}$};
\node[param, below=1.5cm of SD, xshift=1cm, label=right:\small N] (BD) {$\boldsymbol{\beta}_\text{D}$};
\node[param, below=1.4cm of TPHI, xshift=-1cm, label=right:\small CAR] (PK) {$\boldsymbol{\phi}_k$};
\node[param, below=1.5cm of AKAP, xshift=1.5cm, label=right:\small G] (KK) {$\kappa^2_k$};

\node[param, below=7cm of X] (Z) {$\boldsymbol{\zeta}_{jk}$}; 
\node[param, below=7cm of P, label=right:\small IG] (KJK) {$\kappa^2_{jk}$};

\node[obs, below=2cm of Z,xshift=3cm, label=left:\small N] (YIJK) {$y_{ijk}$};


\draw[arrow] (A) -- (SBETA); \draw[arrow] (B) -- (SBETA); 
\draw[arrow] (C) -- (BETA);  \draw[arrow] (SBETA) -- (BETA);

\draw[arrow] (D) -- (SE); \draw[arrow] (E) -- (SE);
\draw[arrow] (F) -- (BE); \draw[arrow] (SE) -- (BE);
\draw[arrow] (G) -- (SM); \draw[arrow] (H) -- (SM);
\draw[arrow] (I) -- (BM); \draw[arrow] (SM) -- (BM);

\draw[arrow] (X) -- (Z);
\draw[arrow] (J) -- (SD); \draw[arrow] (K) -- (SD);
\draw[arrow] (L) -- (BD); \draw[arrow] (SD) -- (BD);

\draw[arrow] (M) -- (PK);
\draw[arrow] (O) -- (TPHI); \draw[arrow] (N) -- (TPHI); \draw[arrow] (TPHI) -- (PK);

\draw[arrow] (P) -- (KJK);
\draw[arrow] (Q) -- (AKAP); \draw[arrow] (R) -- (AKAP);
\draw[arrow] (S) -- (BKAP); \draw[arrow] (T) -- (BKAP);
\draw[arrow] (AKAP) -- (KK); \draw[arrow] (BKAP) -- (KK); \draw[arrow] (KK) -- (KJK);

\draw[arrow] (BETA) -- (Z);
\draw[arrow] (BE) -- (Z);
\draw[arrow] (BM) -- (Z);
\draw[arrow] (BD) -- (Z);
\draw[arrow] (PK) -- (Z);

\draw[arrow] (Z) -- (YIJK); 
\draw[arrow] (KJK) -- (YIJK);

\end{tikzpicture}
}

\caption{Directed acyclic graph (DAG) of the hierarchical model. The observed response is shown in a gray circle, model parameters are shown as circles, and hyperparameters are shown as rectangles. Directed arrows indicate conditional dependence.}
\label{fig:dag_modelo1}
\end{figure}
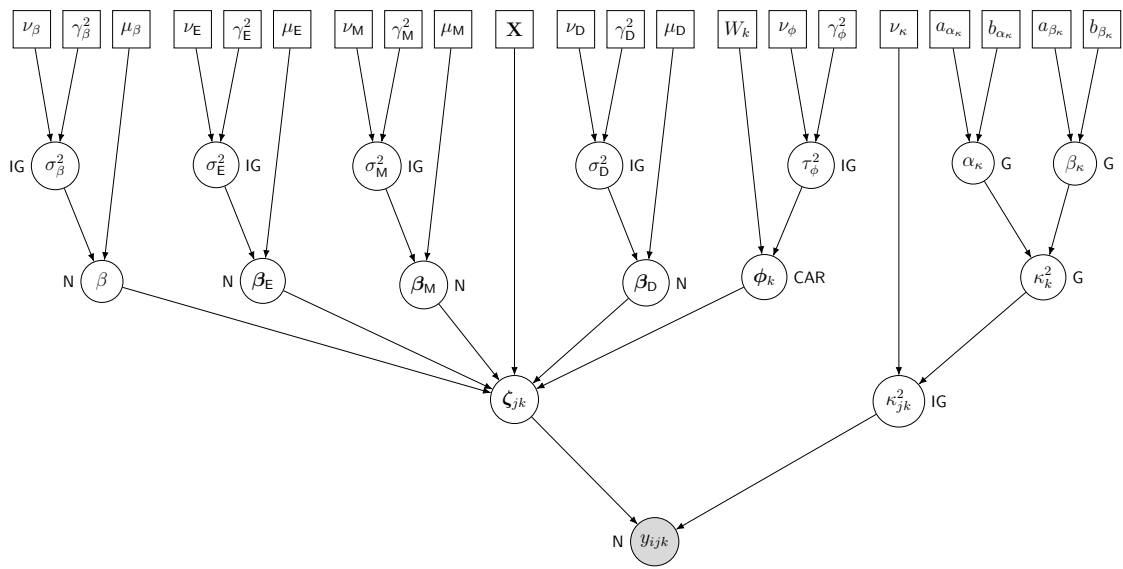

We now assign the following prior distributions
\begin{align*}
\beta_0 &\sim \textsf{N}\!\left(\mu_{\beta_0}, \sigma_{\beta_0}^2\right),
&
\sigma_{\beta_0}^2 &\sim \textsf{IG}\!\left(\frac{\nu_{\beta_0}}{2}, \frac{\nu_{\beta_0}\gamma_{\beta_0}^2}{2}\right),
\\
\boldsymbol{\beta}_{\mathrm{E}} &\sim \textsf{N}_{p_{\mathrm{E}}}\!\left(\boldsymbol{\mu}_{\mathrm{E}}, \sigma_{\mathrm{E}}^2 \mathbf{I}\right),
&
\sigma_{\mathrm{E}}^2 &\sim \textsf{IG}\!\left(\frac{\nu_{\mathrm{E}}}{2}, \frac{\nu_{\mathrm{E}}\gamma_{\mathrm{E}}^2}{2}\right),
\\
\boldsymbol{\beta}_{\mathrm{M}} &\sim \textsf{N}_{p_{\mathrm{M}}}\!\left(\boldsymbol{\mu}_{\mathrm{M}}, \sigma_{\mathrm{M}}^2 \mathbf{I}\right),
&
\sigma_{\mathrm{M}}^2 &\sim \textsf{IG}\!\left(\frac{\nu_{\mathrm{M}}}{2}, \frac{\nu_{\mathrm{M}}\gamma_{\mathrm{M}}^2}{2}\right),
\\
\boldsymbol{\beta}_{\mathrm{D}} &\sim \textsf{N}_{p_{\mathrm{D}}}\!\left(\boldsymbol{\mu}_{\mathrm{D}}, \sigma_{\mathrm{D}}^2 \mathbf{I}\right),
&
\sigma_{\mathrm{D}}^2 &\sim \textsf{IG}\!\left(\frac{\nu_{\mathrm{D}}}{2}, \frac{\nu_{\mathrm{D}}\gamma_{\mathrm{D}}^2}{2}\right),
\\
\kappa_{j,k}^2 &\overset{\text{ind}}{\sim} \textsf{IG}\!\left(\frac{\nu_{\kappa}}{2}, \frac{\nu_{\kappa}\kappa_k^2}{2}\right),
&
\kappa_k^2 &\overset{\text{ind}}{\sim} \textsf{G}\!\left(\frac{\alpha_\kappa}{2}, \frac{\beta_\kappa}{2}\right),
\\
\alpha_\kappa &\sim \textsf{G}\!\left(a_{\alpha_\kappa}, b_{\alpha_\kappa}\right),
&
\beta_\kappa &\sim \textsf{G}\!\left(a_{\beta_\kappa}, b_{\beta_\kappa}\right).
\end{align*}
Normal priors for regression coefficients and inverse gamma priors for variance components yield a conjugate specification under the Gaussian sampling model, which produces tractable full conditional distributions and supports efficient Gibbs updates. 
The gamma priors on $\alpha_\kappa$ and $\beta_\kappa$ provide a flexible and weakly informative layer for the municipal variance hierarchy, ensuring positivity and preserving conditional conjugacy for $\beta_\kappa$, while $\alpha_\kappa$ requires a nonconjugate update.
The municipality level spatial random effects follow an intrinsic conditional autoregressive model,
\begin{align*}
\boldsymbol{\phi}_k \mid \tau_\phi^2, \mathbf{W}_k &\sim \textsf{CAR}\!\left(\tau_\phi^2, \mathbf{W}_k\right),
&
\tau_\phi^2 &\sim \textsf{IG}\!\left(\frac{\nu_\phi}{2}, \frac{\nu_\phi\gamma_\phi^2}{2}\right),
\end{align*}
with the identifiability constraint $\sum_{j=1}^{m_k}\phi_{j,k}=0$ for $k=1,\ldots,d$. This specification is appropriate for areal data because it induces dependence through the adjacency structure and it encourages local smoothing by borrowing strength across neighboring municipalities (see Section \ref{sec_car}). Figure \ref{fig:dag_modelo1} shows the directed acyclic graph (DAG) representation of the hierarchical structure of the model.

\subsection{Normal model with mean and variance parameters and a spatial random effect under Ridge shrinkage}\label{sec_model_ridge}

This specification shares the same likelihood and linear predictor as the baseline model, and it differs only in the prior structure for the regression coefficients. In particular, Ridge shrinkage is introduced through Gaussian priors centered at zero with unknown global precisions, which encourages stable estimation under correlated predictors while retaining the hierarchical and spatial components. Specifically, we retain the sampling model in \eqref{eq_vero_m1} and the mean structure in \eqref{eq_mean_m1}.

\begin{figure}[!htb]
\centering
\resizebox{\textwidth}{!}{%

\begin{tikzpicture}[node distance=0.25cm, font=\huge,
    every node/.style={font=\sffamily\large},
    hyper/.style={draw, rectangle, minimum size=0.8cm, align=center},
    param/.style={draw, circle, minimum size=0.9cm, align=center},
    obs/.style={draw, circle, fill=gray!30, minimum size=1.0cm, align=center},
    arrow/.style={-Latex}]

\node[draw, rectangle, minimum size=1cm] (A) {$\nu_\beta$};
\node[draw, rectangle, minimum size=1cm, right=of A] (B) {$\gamma_\beta^2$};
\node[draw, rectangle, minimum size=1cm, right= of B] (C) {$\mu_\beta$};
\node[draw, rectangle, minimum size=1cm, right= 0.5cm of C] (D) {$\nu_\text{E}$};
\node[draw, rectangle, minimum size=1cm, right= of D] (E) {$\gamma^2_\text{E}$}; 
\node[draw, rectangle, minimum size=1cm, right= 0.5cm of E] (G) {$\nu_\text{M}$};
\node[draw, rectangle, minimum size=1cm, right=of G] (H) {$\gamma^2_\text{M}$};
\node[draw, rectangle, minimum size=1cm, right= 0.5cm of H] (X) {$\mathbf{X}$};
\node[draw, rectangle, minimum size=1cm, right= 0.5cm of X] (J) {$\nu_\text{D}$};
\node[draw, rectangle, minimum size=1cm, right=of J] (K) {$\gamma^2_\text{D}$};
\node[draw, rectangle, minimum size=1cm, right= 0.5cm of K] (M) {$W_k$};
\node[draw, rectangle, minimum size=1cm, right=of M] (N) {$\nu_\phi$};
\node[draw, rectangle, minimum size=1cm, right=of N] (O) {$\gamma^2_\phi$};
\node[draw, rectangle, minimum size=1cm, right=0.5cm of O] (P) {$\nu_\kappa$};
\node[draw, rectangle, minimum size=1cm, right= of P] (Q) {$a_{\alpha_\kappa}$};
\node[draw, rectangle, minimum size=1cm, right= of Q] (R) {$b_{\alpha_\kappa}$};
\node[draw, rectangle, minimum size=1cm, right= of R] (S) {$a_{\beta_\kappa}$};
\node[draw, rectangle, minimum size=1cm, right= of S] (T) {$b_{\beta_\kappa}$};

\node[draw, circle, minimum size=1cm, below=2cm of A, xshift=1cm, label=left:IG] (SBETA) {$\sigma^2_\beta$};
\node[draw, circle, minimum size=1cm, below=2cm of D, xshift=1cm, label=right:G] (SE) {$\lambda^2_\text{E}$};
\node[draw, circle, minimum size=1cm, below=2cm of G, xshift=1cm, label=right:G] (SM) {$\lambda^2_\text{M}$};
\node[draw, circle, minimum size=1cm, below=2cm of J, xshift=1cm, label=right:G] (SD) {$\lambda^2_\text{D}$};
\node[draw, circle, minimum size=1cm, below=2cm of N, xshift=1cm, label=right:IG] (TPHI) {$\tau^2_\phi$};
\node[draw, circle, minimum size=1cm, below=2cm of Q, xshift=1cm, label=right:G] (AKAP) {$\alpha_\kappa$};
\node[draw, circle, minimum size=1cm, below=2cm of S, xshift=1cm, label=right:G] (BKAP) {$\beta_\kappa$};

\node[draw, circle, minimum size=1cm, below=2cm of SBETA, xshift=1cm, label=left:N] (BETA) {$\beta$};
\node[draw, circle, minimum size=1cm, below=2cm of SE, label=left:N] (BE) {$\boldsymbol{\beta}_\text{E}$};
\node[draw, circle, minimum size=1cm, below=2cm of SM, label=right:N] (BM) {$\boldsymbol{\beta}_\text{M}$};
\node[draw, circle, minimum size=1cm, below=2 cm of SD, label=right:N] (BD) {$\boldsymbol{\beta}_\text{D}$};
\node[draw, circle, minimum size=1cm, below=2cm of TPHI, xshift=-1cm, label=right:CAR] (PK) {$\boldsymbol{\phi}_k$};
\node[draw, circle, minimum size=1cm, below=2cm of AKAP, xshift=1.5cm, label=right:G] (KK) {$\kappa^2_k$};

\node[draw, circle, minimum size=1cm, below=8 cm of X,xshift=-1cm] (Z) {$\boldsymbol{\zeta}_{jk}$};
\node[draw, circle, minimum size=1cm, below=8cm of P,xshift=-1cm, label=right:IG] (KJK) {$\kappa^2_{jk}$};

\node[draw, circle, fill=gray!30, minimum size=1.5cm, below=4cm of BD, label=left:N] (YIJK) {$y_{ijk}$};

\draw[->] (A) -- (SBETA); \draw[->] (B) -- (SBETA);
\draw[->] (C) -- (BETA);  \draw[->] (SBETA) -- (BETA);
\draw[->] (D) -- (SE);    \draw[->] (E) -- (SE);
\draw[->] (SE) -- (BE);
\draw[->] (G) -- (SM);    \draw[->] (H) -- (SM);
\draw[->] (SM) -- (BM);
\draw[->] (X) -- (Z);
\draw[->] (J) -- (SD);    \draw[->] (K) -- (SD);
\draw[->] (SD) -- (BD);
\draw[->] (M) -- (PK);
\draw[->] (O) -- (TPHI);  \draw[->] (N) -- (TPHI);
\draw[->] (TPHI) -- (PK);
\draw[->] (P) -- (KJK);
\draw[->] (Q) -- (AKAP);  \draw[->] (R) -- (AKAP);
\draw[->] (S) -- (BKAP);  \draw[->] (T) -- (BKAP);
\draw[->] (AKAP) -- (KK); \draw[->] (BKAP) -- (KK);
\draw[->] (KK) -- (KJK);
\draw[->] (BETA) -- (Z);  \draw[->] (BE) -- (Z);
\draw[->] (BM) -- (Z);    \draw[->] (BD) -- (Z);
\draw[->] (PK) -- (Z);
\draw[->] (Z) -- (YIJK);  \draw[->] (KJK) -- (YIJK); 

\end{tikzpicture}
}

\caption{Directed acyclic graph (DAG) visualizing the hierarchical model under Ridge shrinkage.}
\label{fig:dag_modelo_ridge}
\end{figure}
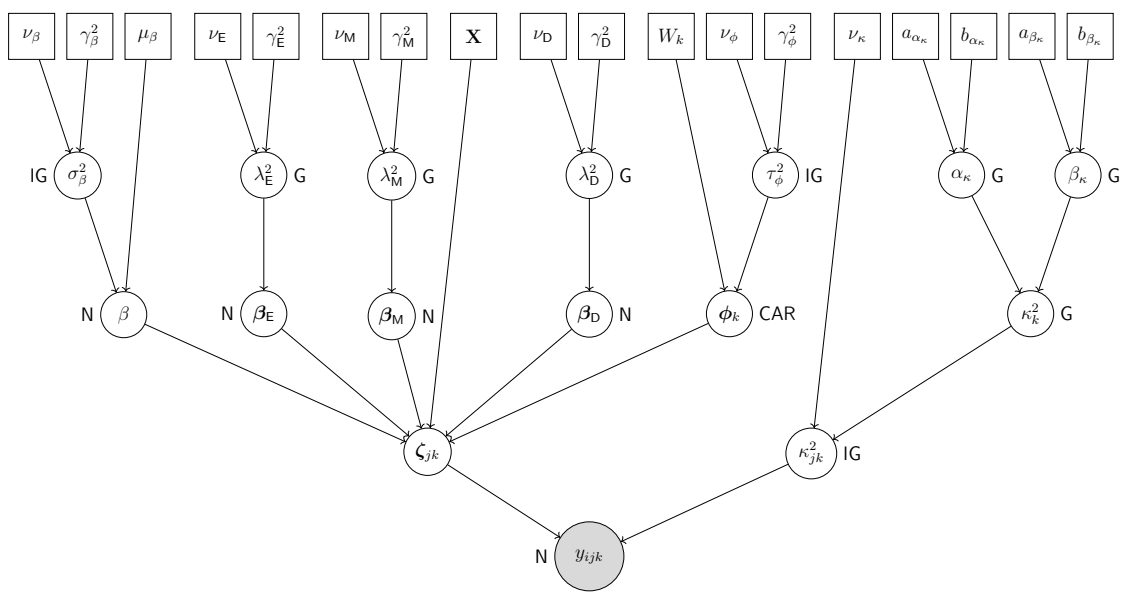

All prior distributions are as in the baseline model in Section \ref{sec_model1}, except for the regression coefficients, which now follow Ridge shrinkage priors (see Section \ref{sec_regu}). Specifically,
\begin{align*}
\boldsymbol{\beta}_{\mathrm{E}} \mid \lambda_{\mathrm{E}}^2
&\sim \textsf{N}_{p_{\mathrm{E}}}\!\left(\mathbf{0}, \frac{1}{\lambda_{\mathrm{E}}^2}\mathbf{I}\right),
&
\lambda_{\mathrm{E}}^2
&\sim \textsf{G}\!\left(\frac{\nu_{\mathrm{E}}}{2}, \frac{\nu_{\mathrm{E}}\gamma_{\mathrm{E}}^2}{2}\right),
\\
\boldsymbol{\beta}_{\mathrm{M}} \mid \lambda_{\mathrm{M}}^2
&\sim \textsf{N}_{p_{\mathrm{M}}}\!\left(\mathbf{0}, \frac{1}{\lambda_{\mathrm{M}}^2}\mathbf{I}\right),
&
\lambda_{\mathrm{M}}^2
&\sim \textsf{G}\!\left(\frac{\nu_{\mathrm{M}}}{2}, \frac{\nu_{\mathrm{M}}\gamma_{\mathrm{M}}^2}{2}\right),
\\
\boldsymbol{\beta}_{\mathrm{D}} \mid \lambda_{\mathrm{D}}^2
&\sim \textsf{N}_{p_{\mathrm{D}}}\!\left(\mathbf{0}, \frac{1}{\lambda_{\mathrm{D}}^2}\mathbf{I}\right),
&
\lambda_{\mathrm{D}}^2
&\sim \textsf{G}\!\left(\frac{\nu_{\mathrm{D}}}{2}, \frac{\nu_{\mathrm{D}}\gamma_{\mathrm{D}}^2}{2}\right).
\end{align*}
This Ridge specification induces global shrinkage toward zero through $(\lambda_{\mathrm{E}}^2,\lambda_{\mathrm{M}}^2,\lambda_{\mathrm{D}}^2)$, which stabilizes inference under multicollinearity while preserving the multilevel and spatial structure of the baseline model from Section \ref{sec_model1}, including the municipality variance hierarchy and the intrinsic CAR prior for the spatial random effects.
Figure \ref{fig:dag_modelo_ridge} shows the DAG representation of the hierarchical structure of the model.

\subsection{Normal model with mean and variance parameters and a spatial random effect under Lasso shrinkage}

This specification shares the same likelihood and linear predictor as the baseline model in Section \ref{sec_model1} and the Ridge model in Section \ref{sec_model_ridge}, and it differs only in the prior structure assigned to the regression coefficients. Here, Lasso shrinkage is introduced by replacing the Ridge Gaussian priors with Laplace type priors, which induce stronger shrinkage toward zero and may yield sparse posterior summaries by concentrating weak effects near zero.

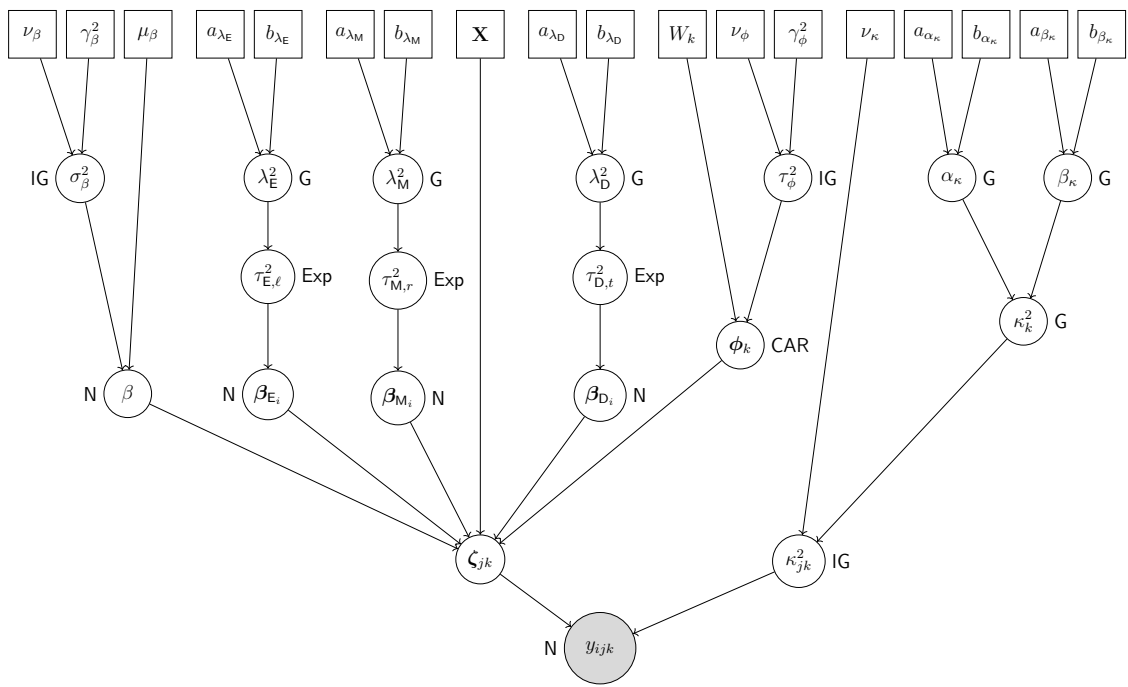
\begin{figure}[!htb]
\centering
\resizebox{\textwidth}{!}{%
\begin{tikzpicture}[node distance=0.2cm,font=\huge,
    every node/.style={font=\sffamily\large},
    hyper/.style={draw, rectangle, minimum size=0.8cm, align=center},
    param/.style={draw, circle, minimum size=0.9cm, align=center},
    obs/.style={draw, circle, fill=gray!30, minimum size=1.0cm, align=center},
    arrow/.style={-Latex}]

\node[draw, rectangle, minimum size=1cm] (A) {$\nu_\beta$};
\node[draw, rectangle, minimum size=1cm, right=of A] (B) {$\gamma_\beta^2$};
\node[draw, rectangle, minimum size=1cm, right= of B] (C) {$\mu_\beta$};
\node[draw, rectangle, minimum size=1cm, right= 0.5cm of C] (D) {$a_{\lambda_\text{E}}$};
\node[draw, rectangle, minimum size=1cm, right= of D] (E) {$b_{\lambda_\text{E}}$};
\node[draw, rectangle, minimum size=1cm, right= 0.5cm of E] (G) {$a_{\lambda_\text{M}}$};
\node[draw, rectangle, minimum size=1cm, right=of G] (H) {$b_{\lambda_\text{M}}$};
\node[draw, rectangle, minimum size=1cm, right= 0.5cm of H] (X) {$\mathbf{X}$};
\node[draw, rectangle, minimum size=1cm, right= 0.5cm of X] (J) {$a_{\lambda_\text{D}}$};
\node[draw, rectangle, minimum size=1cm, right=of J] (K) {$b_{\lambda_\text{D}}$};
\node[draw, rectangle, minimum size=1cm, right= 0.5cm of K] (M) {$W_k$};
\node[draw, rectangle, minimum size=1cm, right=of M] (N) {$\nu_\phi$};
\node[draw, rectangle, minimum size=1cm, right=of N] (O) {$\gamma^2_\phi$};
\node[draw, rectangle, minimum size=1cm, right=0.5cm of O] (P) {$\nu_\kappa$};
\node[draw, rectangle, minimum size=1cm, right= of P] (Q) {$a_{\alpha_\kappa}$};
\node[draw, rectangle, minimum size=1cm, right= of Q] (R) {$b_{\alpha_\kappa}$};
\node[draw, rectangle, minimum size=1cm, right= of R] (S) {$a_{\beta_\kappa}$};
\node[draw, rectangle, minimum size=1cm, right= of S] (T) {$b_{\beta_\kappa}$};

\node[draw, circle, minimum size=1cm, below=2cm of A, xshift=1cm, label=left:IG] (SBETA) {$\sigma^2_\beta$};
\node[draw, circle, minimum size=1cm, below=2cm of D, xshift=1cm, label=right:G] (SE) {$\lambda^2_\text{E}$};
\node[draw, circle, minimum size=1cm, below=2cm of G, xshift=1cm, label=right:G] (SM) {$\lambda^2_\text{M}$};
\node[draw, circle, minimum size=1cm, below=2cm of J, xshift=1cm, label=right:G] (SD) {$\lambda^2_\text{D}$};
\node[draw, circle, minimum size=1cm, below=2cm of N, xshift=1cm, label=right:IG] (TPHI) {$\tau^2_\phi$};
\node[draw, circle, minimum size=1cm, below=2cm of Q, xshift=0.5cm, label=right:G] (AKAP) {$\alpha_\kappa$};
\node[draw, circle, minimum size=1cm, below=2cm of S, xshift=0.5cm, label=right:G] (BKAP) {$\beta_\kappa$};

\node[draw, circle, minimum size=1cm, below=1cm of SE, label=right:Exp] (TE) {$\tau^2_{\text{E},\ell}$};
\node[draw, circle, minimum size=1cm, below=1cm of SM, label=right:Exp] (TM) {$\tau^2_{\text{M},r}$};
\node[draw, circle, minimum size=1cm, below=1cm of SD, label=right:Exp] (TD) {$\tau^2_{\text{D},t}$};

\node[draw, circle, minimum size=1cm, below=3.5cm of SBETA, xshift=1cm, label=left:N] (BETA) {$\beta$};
\node[draw, circle, minimum size=1cm, below=3.5cm of SE, label=left:N] (BE) {$\boldsymbol{\beta}_{\text{E}_i}$};
\node[draw, circle, minimum size=1cm, below=3.5cm of SM, label=right:N] (BM) {$\boldsymbol{\beta}_{\text{M}_i}$};
\node[draw, circle, minimum size=1cm, below=3.5cm of SD, label=right:N] (BD) {$\boldsymbol{\beta}_{\text{D}_i}$}; 
\node[draw, circle, minimum size=1cm, below=2.5cm of TPHI, xshift=-1cm, label=right:CAR] (PK) {$\boldsymbol{\phi}_k$};
\node[draw, circle, minimum size=1cm, below=2cm of AKAP, xshift=1.5cm, label=right:G] (KK) {$\kappa^2_k$};

\node[draw, circle, minimum size=1cm, below=10cm of X] (Z) {$\boldsymbol{\zeta}_{jk}$};
\node[draw, circle, minimum size=1cm, below=10cm of P,xshift=-1.5cm, label=right:IG] (KJK) {$\kappa^2_{jk}$};
\node[draw, circle, fill=gray!30,minimum size=1.5cm, below=4cm of BD, label=left:N] (YIJK) {$y_{ijk}$};

\draw[->] (A) -- (SBETA); \draw[->] (B) -- (SBETA); \draw[->] (C) -- (BETA); \draw[->] (SBETA) -- (BETA);

\draw[->] (D) -- (SE); \draw[->] (E) -- (SE); \draw[->] (TE) -- (BE); \draw[->] (SE) -- (TE);
\draw[->] (G) -- (SM); \draw[->] (H) -- (SM); \draw[->] (TM) -- (BM); \draw[->] (SM) -- (TM);
\draw[->] (J) -- (SD); \draw[->] (K) -- (SD); \draw[->] (TD) -- (BD); \draw[->] (SD) -- (TD);
\draw[->] (X) -- (Z);

\draw[->] (M) -- (PK);
\draw[->] (O) -- (TPHI); \draw[->] (N) -- (TPHI); 
\draw[->] (TPHI) -- (PK); 

\draw[->] (P) -- (KJK);
\draw[->] (Q) -- (AKAP); \draw[->] (R) -- (AKAP); \draw[->] (S) -- (BKAP); \draw[->] (T) -- (BKAP);
\draw[->] (AKAP) -- (KK); \draw[->] (BKAP) -- (KK); \draw[->] (KK) -- (KJK);

\draw[->] (BETA) -- (Z); \draw[->] (BE) -- (Z); \draw[->] (BM) -- (Z); \draw[->] (BD) -- (Z);
\draw[->] (PK) -- (Z);
\draw[->] (Z) -- (YIJK); \draw[->] (KJK) -- (YIJK); 

\end{tikzpicture}
}

\caption{Directed acyclic graph (DAG) visualizing the hierarchical model under Lasso shrinkage.}
\label{fig:dag_modelo_lasso}
\end{figure}

We again retain the sampling model in \eqref{eq_vero_m1} with the same mean structure in \eqref{eq_mean_m1}. All prior distributions are as in the baseline model, except for the regression coefficients, which now follow Bayesian Lasso priors (see Section \ref{sec_regu}). To facilitate posterior computation, we adopt the normal exponential scale mixture representation of the Laplace distribution, so that, for $\ell=1,\ldots,p_{\mathrm{E}}$, $r=1,\ldots,p_{\mathrm{M}}$, and $t=1,\ldots,p_{\mathrm{D}}$,
\begin{align*}
\beta_{\mathrm{E},\ell} \mid \tau_{\mathrm{E},\ell}^2
&\overset{\text{ind}}{\sim}
\textsf{N}\!\left(0,\tau_{\mathrm{E},\ell}^2\right),
&
\tau_{\mathrm{E},\ell}^2 \mid \lambda_{\mathrm{E}}^2
&\overset{\text{ind}}{\sim}
\textsf{Exp}\!\left(\frac{\lambda_{\mathrm{E}}^2}{2}\right),
&
\lambda_{\mathrm{E}}^2
&\sim
\textsf{G}\!\left(a_{\lambda_{\mathrm{E}}},b_{\lambda_{\mathrm{E}}}\right),
\\
\beta_{\mathrm{M},r} \mid \tau_{\mathrm{M},r}^2
&\overset{\text{ind}}{\sim}
\textsf{N}\!\left(0,\tau_{\mathrm{M},r}^2\right),
&
\tau_{\mathrm{M},r}^2 \mid \lambda_{\mathrm{M}}^2
&\overset{\text{ind}}{\sim}
\textsf{Exp}\!\left(\frac{\lambda_{\mathrm{M}}^2}{2}\right),
&
\lambda_{\mathrm{M}}^2
&\sim
\textsf{G}\!\left(a_{\lambda_{\mathrm{M}}},b_{\lambda_{\mathrm{M}}}\right),
\\
\beta_{\mathrm{D},t} \mid \tau_{\mathrm{D},t}^2
&\overset{\text{ind}}{\sim}
\textsf{N}\!\left(0,\tau_{\mathrm{D},t}^2\right),
&
\tau_{\mathrm{D},t}^2 \mid \lambda_{\mathrm{D}}^2
&\overset{\text{ind}}{\sim}
\textsf{Exp}\!\left(\frac{\lambda_{\mathrm{D}}^2}{2}\right),
&
\lambda_{\mathrm{D}}^2
&\sim
\textsf{G}\!\left(a_{\lambda_{\mathrm{D}}},b_{\lambda_{\mathrm{D}}}\right).
\end{align*}
This hierarchy is equivalent to independent Laplace priors on the coefficients and yields conditionally Gaussian updates given the latent scales, which supports efficient posterior computation while allowing coefficient specific shrinkage. The remaining priors are inherited from Section \ref{sec_model1}. 
Figure \ref{fig:dag_modelo_lasso} shows the DAG representation of the hierarchical structure of the model.\\

\subsection{Computation}\label{sec_computation}

For a given model, the posterior distribution of all unknown quantities can be explored using Markov chain Monte Carlo (MCMC) methods (e.g., \citealt{Gamerman2006MCMC}). In particular, we generate a Markov chain $\boldsymbol{\Theta}^{(1)},\ldots,\boldsymbol{\Theta}^{(B)}$ whose stationary distribution is the posterior distribution $p(\mathbf{\Theta}\mid\mathbf{y},\mathbf{X})$, where $\boldsymbol{\Theta}$ collects the regression coefficients, variance components, spatial random effects, shrinkage parameters, and any other model-specific quantities. Posterior point and interval estimates are then obtained from the corresponding empirical distributions. Posterior simulation is carried out via a Gibbs sampler because all full conditional distributions are available in closed form, except for $\alpha_\kappa$, which is updated using a Metropolis step to accommodate its support constraints (e.g., \citealt{Hoff2009}). The complete set of full conditional distributions is reported in Appendix \ref{app_fcds}.

Table \ref{tab_param} summarizes the number of parameters and hyperparameters in each model fitted to the Saber 11 data. The dataset includes $m=1111$ municipalities, $d=32$ departments, and a total sample size of $n=235,601$. The sample size is fixed and can be decomposed by department and by municipality within department as $n=\sum_{k=1}^{d} n_k=\sum_{k=1}^{d}\sum_{j=1}^{m_k} n_{j,k}$, where $n_k$ denotes the number of students in department $k$, $m_k$ is the number of municipalities in department $k$, and $n_{j,k}$ is the number of students in municipality $j$ within department $k$. 
Overall, these models are high dimensional, involving thousands of unknown quantities, which makes the computational strategy implemented in our publicly available GitHub repository particularly valuable. The code is available at {\footnotesize \url{https://github.com/laura-p20/Bayesian-multilevel-model-with-spatial-random-effect}}.

\begin{table}[!htb]
\centering
\begin{tabular}{lcc}
\hline
Model & \# parameters & \# hyperparameters \\
\hline
Baseline & 237,895 & 1,132 \\
Ridge    & 237,895 & 1,129 \\
Lasso    & 525,095 & 1,129 \\
\hline
\end{tabular}
\caption{Number of parameters and hyperparameters in each model.}
\label{tab_param}
\end{table}

Regarding hyperparameter elicitation, we adopt a unit information prior approach \citep{kass1996selection} and center the regression coefficient priors at the corresponding OLS estimates. The remaining hyperparameters are fixed at $\nu_{\phi}=\nu_{\beta_0}=\nu_{\mathrm{E}}=\nu_{\mathrm{M}}=\nu_{\mathrm{D}}=\nu_{\kappa}=2$, $\gamma_{\phi}=\gamma_{\beta_0}=\gamma_{\mathrm{E}}=\gamma_{\mathrm{M}}=\gamma_{\mathrm{D}}=1$, $a_{\alpha_{\kappa}}=a_{\beta_{\kappa}}=2$, and $b_{\alpha_{\kappa}}=b_{\beta_{\kappa}}=5$ to obtain a weakly informative specification that stabilizes variance components and shrinkage parameters while keeping the data contribution to the prior specification as modest as possible.

\section{Saber 11 data revisited}\label{sec_data_analysis}

This section reports posterior inference for sociodemographic effects and territorial segmentation from three complementary perspectives, including departmental and municipal components, and an additional segmentation based on municipality level spatial random effects that captures residual patterns not explained by observed covariates. Posterior summaries for covariate effects are presented for all three model variants, whereas the hierarchical segmentation is reported using the Ridge model given its strong predictive performance. The segmentation based on spatial random effects is obtained from the baseline model, since it yields spatial effects that remain largely uninfluenced by the covariates, supporting an assessment of both structured and residual territorial variability in educational outcomes.

All results reported below are based on fitting the three model variants via a Gibbs sampler with 127{,}500 iterations per model. The first 10\% of draws were discarded as burn in, and the remaining chain was thinned every five iterations to reduce autocorrelation, yielding 25{,}000 posterior samples for inference. Given the high dimensional parameter space, the log likelihood was recorded at each iteration to monitor sampling behavior, and the resulting traceplots (not shown) indicate stable exploration with no evident signs of nonconvergence. Additional diagnostics, including effective sample sizes and Monte Carlo standard errors for the parameters in each model (not shown), do not reveal problematic values and are consistent with satisfactory convergence. All diagnostics and supporting figures can be replicated using the GitHub repository provided in Section \ref{sec_computation}.

\subsection{Posterior assessment of sociodemographic covariates and territorial segmentation}

\begin{figure}[!htb]
\centering
\includegraphics[width=\linewidth,trim={0cm 0cm 0cm 1.5cm},clip]{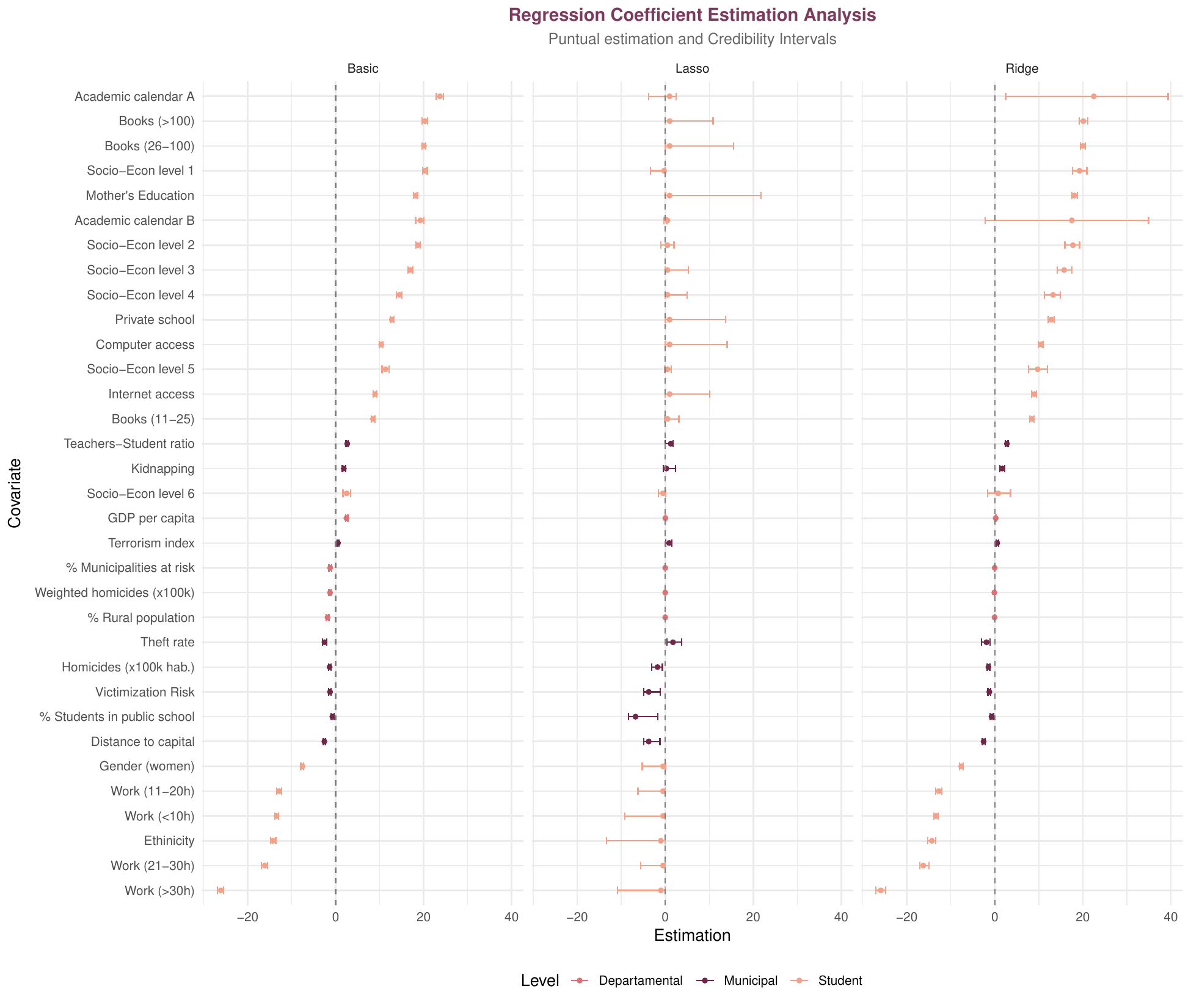}
\caption{Posterior point estimates and 95\% credible intervals for regression coefficients across model variants and hierarchical levels.}
\label{fig:post_covs_all}
\end{figure}

This section reports posterior results from the three model variants to facilitate comparison across shrinkage specifications. Because municipal and departmental covariates are standardized before fitting, regression effects are interpreted on the standardized scale rather than in the original measurement units. Figure \ref{fig:post_covs_all} summarizes posterior point estimates and 95\% credible intervals for sociodemographic covariates across the three hierarchical levels under each model. Overall, covariates capturing students' individual living conditions show the strongest associations with the Saber 11 global score, whereas municipality and department covariates have more moderate contributions. Differences across models align with their shrinkage behavior, most notably under the Lasso specification, which more aggressively contracts smaller effects toward zero.

\subsubsection{Student level covariates}

The student level analysis identifies several covariates with substantial associations with academic performance (see Table \ref{tab_Cov_estu_eval}). Students enrolled in academic calendars A and B show higher scores than those outside these calendars, which may reflect that the latter group is more likely to include students in nontraditional educational modalities, including cycle based programs, flexible schedules linked to work obligations, or delayed schooling. Access to books at home is also associated with higher performance, underscoring the relevance of learning resources. Housing conditions appear similarly important, since students living in stratified dwellings outperform those in non stratified homes, a baseline category that often concentrates severe deprivation and limited access to basic services. Maternal education is another key factor, with higher scores among students whose mothers completed higher education relative to those whose mothers did not attain a college degree. School type also matters, as private school students tend to achieve higher scores than their public school peers. Finally, access to digital resources, including a computer and internet connection, is associated with better performance, highlighting the role of technological support at home.

\begin{table}[!htb]
\centering
\normalsize
\resizebox{\textwidth}{!}{%
\begin{tabular}{p{3.5cm} ccc c  ccc c  ccc c}
\toprule
& \multicolumn{4}{c}{Baseline model}
& \multicolumn{4}{c}{Ridge    model}
& \multicolumn{4}{c}{Lasso    model}
\\
\cmidrule(lr){2-5}\cmidrule(lr){6-9}\cmidrule(lr){10-13}
Variable
& Mean & Lower & Upper & Sig.
& Mean & Lower & Upper & Sig.
& Mean & Lower & Upper & Sig.
\\
\midrule
Mother's education      & 18.15 & 17.65 & 18.59 & 1 & 18.10 & 17.57 & 18.74 & 1 &  1.00 &  0.03 & 21.82 & 1 \\
Computer access         & 10.35 &  9.96 & 10.74 & 1 & 10.45 &  9.93 & 10.91 & 1 &  1.00 &  0.05 & 14.07 & 1 \\
Internet access         &  8.95 &  8.50 &  9.34 & 1 &  8.95 &  8.38 &  9.46 & 1 &  1.00 &  0.03 & 10.11 & 1 \\
Ethnicity               &-14.30 &-14.82 &-13.63 & 1 &-14.30 &-15.19 &-13.40 & 1 & -1.00 &-13.36 & -0.01 & 1 \\
Books (11--25)          &  8.45 &  8.10 &  8.86 & 1 &  8.45 &  7.98 &  8.88 & 1 &  0.50 & -0.11 &  3.10 & 0 \\
Books (26--100)         & 20.05 & 19.59 & 20.50 & 1 & 20.10 & 19.41 & 20.59 & 1 &  1.00 &  0.03 & 15.55 & 1 \\
Books ($>$100)          & 20.30 & 19.61 & 20.91 & 1 & 20.10 & 19.15 & 21.12 & 1 &  1.00 &  0.00 & 10.87 & 1 \\
Socio-Econ level 1      & 20.30 & 19.81 & 20.81 & 1 & 19.25 & 17.62 & 20.90 & 1 & -0.25 & -3.34 &  0.02 & 0 \\
Socio-Econ level 2      & 18.70 & 18.20 & 19.18 & 1 & 17.75 & 15.92 & 19.24 & 1 &  0.50 & -0.99 &  2.00 & 0 \\
Socio-Econ level 3      & 16.90 & 16.43 & 17.48 & 1 & 15.75 & 14.16 & 17.51 & 1 &  0.50 & -0.01 &  5.24 & 0 \\
Socio-Econ level 4      & 14.50 & 13.77 & 15.07 & 1 & 13.25 & 11.28 & 14.88 & 1 &  0.50 & -0.01 &  5.00 & 0 \\
Socio-Econ level 5      & 11.30 & 10.54 & 12.12 & 1 &  9.75 &  7.63 & 11.92 & 1 &  0.50 & -0.11 &  1.36 & 0 \\
Socio-Econ level 6      &  2.50 &  1.65 &  3.37 & 1 &  0.75 & -1.69 &  3.57 & 0 & -0.50 & -1.60 &  0.09 & 0 \\
Gender (women)          & -7.55 & -7.93 & -7.25 & 1 & -7.65 & -8.04 & -7.24 & 1 & -0.50 & -5.22 &  0.05 & 0 \\
Academic calendar A     & 23.70 & 22.90 & 24.50 & 1 & 22.50 &  2.45 & 39.39 & 1 &  1.00 & -3.73 &  2.50 & 0 \\
Academic calendar B     & 19.25 & 18.16 & 20.00 & 1 & 17.50 & -2.20 & 34.94 & 0 &  0.50 & -0.28 &  0.31 & 0 \\
Private school          & 12.75 & 12.36 & 13.25 & 1 & 12.90 & 12.16 & 13.51 & 1 &  1.00 &  0.03 & 13.73 & 1 \\
Work ($<$10h)           &-13.45 &-13.84 &-13.03 & 1 &-13.35 &-13.88 &-12.90 & 1 & -0.50 & -9.16 & -0.02 & 1 \\
Work (11--20h)          &-12.85 &-13.38 &-12.34 & 1 &-12.70 &-13.44 &-12.08 & 1 & -0.50 & -6.21 &  0.00 & 0 \\
Work (21--30h)          &-16.10 &-16.86 &-15.48 & 1 &-16.25 &-17.10 &-14.94 & 1 & -0.50 & -5.59 &  0.02 & 0 \\
Work ($>$30h)           &-26.10 &-26.82 &-25.45 & 1 &-25.90 &-27.00 &-24.82 & 1 & -1.00 &-10.87 &  0.00 & 0 \\
\bottomrule
\end{tabular}%
}
\caption{Student level regression coefficient posterior summaries under the baseline, Ridge, and Lasso specifications. Lower and Upper denote the lower and upper limits of the 95\% credible interval. Sig.\ equals 1 if the interval excludes zero and 0 otherwise.}
\label{tab_Cov_estu_eval}
\end{table}

Conversely, several student level factors are associated with significant disadvantages relative to the baseline categories. Being female is linked to lower expected scores than being male, which is consistent with persistent gender gaps in educational outcomes. Working while studying is also associated with lower expected scores relative to not working, suggesting that time and resource constraints can hinder academic performance. Ethnicity is another salient factor, as students belonging to minority ethnic groups exhibit reduced global scores compared to the reference category. Taken together, these results point to deep and intersecting inequalities, since disadvantages may compound for students who simultaneously face multiple constraints, for example women from minority ethnic groups who also work while studying. Overall, the posterior summaries indicate substantial disparities that reflect structural differences in opportunities and learning conditions.

From a model comparison perspective, all student level covariates are significant under the baseline specification. Under Ridge shrinkage, the only student level covariates that are not significant are Socio Economic Level 6 and Academic Calendar B, while the remaining effects retain credible intervals that exclude zero. Under Lasso shrinkage, the set of selected covariates is smaller and concentrates on the strongest signals, including Mother’s Education, computer access, internet access, ethnicity, all categories of books at home, private school attendance, and working less than 10 hours per week. This pattern is consistent with the more aggressive shrinkage of Lasso, which shrinks weaker effects toward zero while retaining covariates with the largest associations with the Saber 11 global score.

\subsubsection{Municipal level covariates}

As shown in Tables \ref{tab_mun_cov} and \ref{tab_dep_cov}, municipal and departmental covariates exhibit smaller marginal associations with the global score than student level characteristics. Among municipal covariates, the teacher to student ratio shows a positive association, which suggests that higher teacher availability is linked to slightly higher scores. In contrast, the terrorism measure has an almost negligible effect. Higher theft and homicide rates, higher victimization risk, a larger share of students enrolled in public schools, and greater distance to the departmental capital are associated with lower global scores, with effects that are modest but non negligible.

\begin{table}[!htb]
\centering
\normalsize
\resizebox{\textwidth}{!}{%
\begin{tabular}{p{3.5cm} ccc c  ccc c  ccc c}
\toprule
& \multicolumn{4}{c}{Baseline model}
& \multicolumn{4}{c}{Ridge model}
& \multicolumn{4}{c}{Lasso model}
\\
\cmidrule(lr){2-5}\cmidrule(lr){6-9}\cmidrule(lr){10-13}
Variable
& Mean & Lower & Upper & Sig.
& Mean & Lower & Upper & Sig.
& Mean & Lower & Upper & Sig.
\\
\midrule
Teacher to student ratio           &  2.63 &  2.35 &  2.86 & 1 &  2.75 &  2.43 &  3.02 & 1 &  1.25 & -0.05 &  1.81 & 0 \\
Victimization risk                 & -1.28 & -1.56 & -1.03 & 1 & -1.25 & -1.57 & -0.98 & 1 & -3.75 & -4.91 & -1.16 & 1 \\
Homicides per 100{,}000 inhabitants& -1.38 & -1.59 & -1.11 & 1 & -1.45 & -1.70 & -1.15 & 1 & -1.75 & -3.04 & -0.64 & 1 \\
Students in public school (\%)     & -0.75 & -1.00 & -0.41 & 1 & -0.75 & -1.10 & -0.33 & 1 & -6.75 & -8.38 & -1.72 & 1 \\
Terrorism index                    &  0.53 &  0.30 &  0.77 & 1 &  0.55 &  0.32 &  0.86 & 1 &  0.90 &  0.06 &  1.46 & 1 \\
Theft rate                         & -2.55 & -3.01 & -2.04 & 1 & -1.90 & -3.00 & -1.08 & 1 &  1.75 &  0.43 &  3.69 & 1 \\
Kidnapping                         &  1.85 &  1.52 &  2.24 & 1 &  1.70 &  1.12 &  2.29 & 1 &  0.25 & -0.41 &  2.30 & 0 \\
Distance to capital                & -2.58 & -2.80 & -2.33 & 1 & -2.55 & -2.82 & -2.23 & 1 & -3.75 & -4.92 & -1.21 & 1 \\
\bottomrule
\end{tabular}%
}
\caption{Municipal level regression coefficient posterior summaries under the baseline, Ridge, and Lasso specifications. Lower and Upper denote the lower and upper limits of the 95\% credible interval. Sig.\ equals 1 if the interval excludes zero and 0 otherwise.}
\label{tab_mun_cov}
\end{table}

\subsubsection{Departmental level covariates}

At the departmental level, GDP per capita is associated with a small positive effect, whereas the share of municipalities at risk, the weighted homicide rate, and the rural population proportion show small negative effects. Overall, these results suggest that regional characteristics contribute to performance, but their influence is considerably smaller than that of individual student conditions. In the baseline and Ridge specifications, all departmental and municipal covariates are significant. In contrast, the Lasso specification does not select the departmental covariates, whereas it retains several municipal covariates, which suggests that departmental information is largely captured by municipal level variation.

\begin{table}[!htb]
\centering
\normalsize
\resizebox{\textwidth}{!}{%
\begin{tabular}{p{3.5cm} ccc c  ccc c  ccc c}
\toprule
& \multicolumn{4}{c}{Baseline model}
& \multicolumn{4}{c}{Ridge model}
& \multicolumn{4}{c}{Lasso model}
\\
\cmidrule(lr){2-5}\cmidrule(lr){6-9}\cmidrule(lr){10-13}
Variable
& Mean & Lower & Upper & Sig.
& Mean & Lower & Upper & Sig.
& Mean & Lower & Upper & Sig.
\\
\midrule
GDP per capita                     &  2.45 &  2.23 &  2.77 & 1 &  0.20 &  0.17 &  0.22 & 1 &  0.03 & -0.00 &  0.03 & 0 \\
Rural population (\%)              & -1.85 & -2.21 & -1.54 & 1 & -0.07 & -0.08 & -0.05 & 1 & -0.01 & -0.04 &  0.00 & 0 \\
Municipalities at risk (\%)        & -1.25 & -1.55 & -0.89 & 1 & -0.04 & -0.05 & -0.02 & 1 & -0.01 & -0.01 &  0.00 & 0 \\
Weighted homicides per 100{,}000    & -1.28 & -1.55 & -1.04 & 1 & -0.11 & -0.13 & -0.08 & 1 & -0.01 & -0.01 &  0.00 & 0 \\
\bottomrule
\end{tabular}%
}
\caption{Departmental level regression coefficient posterior summaries under the baseline, Ridge, and Lasso specifications. Lower and Upper denote the lower and upper limits of the 95\% credible interval. Sig.\ equals 1 if the interval excludes zero and 0 otherwise.}
\label{tab_dep_cov}
\end{table}

\subsection{Spatial random effects}

\begin{figure}[!htb]
    \centering
    \includegraphics[width=0.65\linewidth,trim={2cm 0cm 1cm 1.5cm},clip]{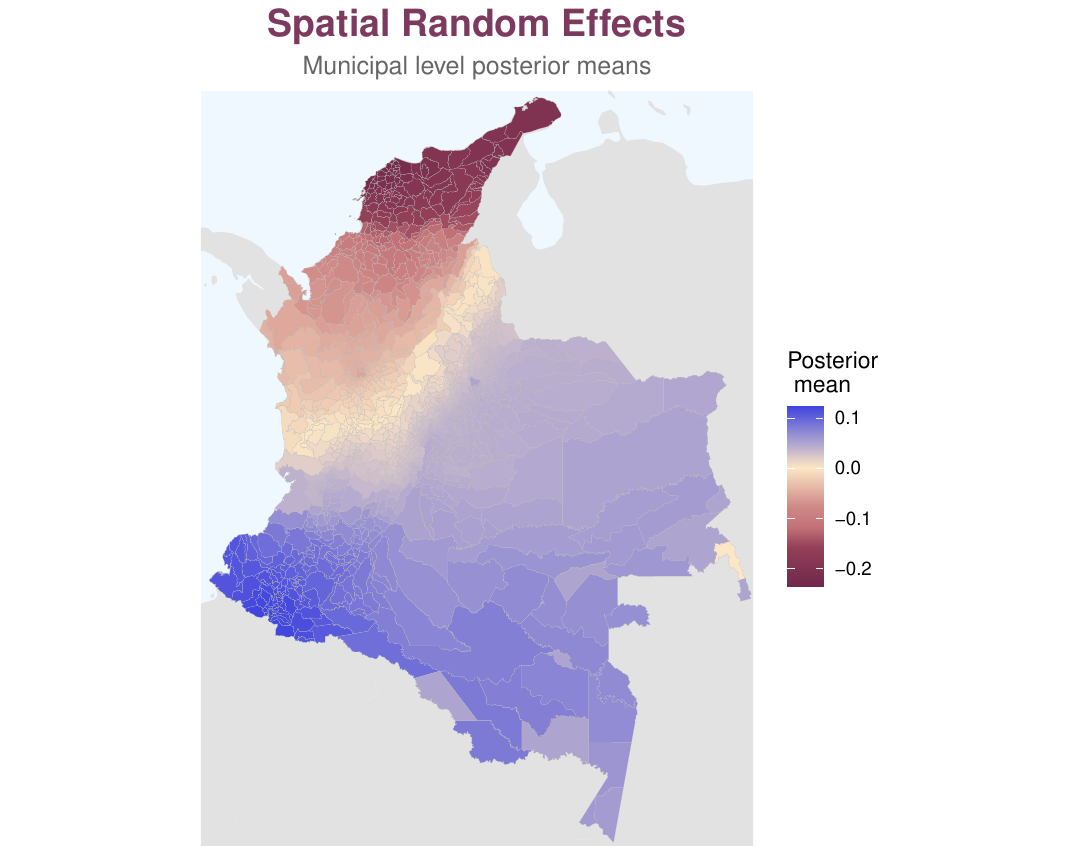}
    \caption{Posterior mean of the municipal spatial random effects.}
    \label{fig:spatial_map}
\end{figure}

Here we examine the municipal random effects in $\boldsymbol{\phi}_k$ by computing their posterior means. We use the baseline specification given in Section \ref{sec_model1} because it allows the spatial random effects to remain largely uninfluenced by covariates. Figure \ref{fig:spatial_map} shows the posterior mean of the municipal spatial random effects. Interpretation requires caution because these effects are subject to identifiability constraints, so they are not directly interpretable on the global score scale. However, their sign and spatial gradients remain informative. The posterior means suggest a negative spatial component in the northern region of the country, which indicates residual spatial influence not explained by the covariates. This negative effect weakens toward the central region, where posterior means are close to zero, which indicates no additional spatial contribution beyond the covariate effects. South of this area, a mild positive spatial effect appears and becomes more pronounced in the southwestern region, where location is associated with stronger positive deviations. These patterns should be interpreted as departures from the covariate explained component rather than as marginal effects of geography on the observed score.

\subsection{Departmental segmentation based on mean global scores}\label{sec_dep_seg}

Here we use posterior draws of the departmental means, obtained by averaging the student specific means $\zeta_{i,j,k}$ within each department, to construct a departmental segmentation. At each MCMC iteration $b$, we fit a k means algorithm and select the number of clusters using the silhouette criterion \citep{james2013islr}. Given the resulting partition, we define a binary adjacency matrix $\mathbf{A}^{(b)}$, with entries $A_{i,j}^{(b)}=1$ if departments $i$ and $j$ belong to the same cluster and $0$ otherwise. The posterior co-clustering probability matrix is $\mathbf{P}=\textsf{E}(\mathbf{A} \mid \mathbf{y})$, which we approximate by $\mathbf{P}=\frac{1}{B}\sum_{b=1}^{B}\mathbf{A}^{(b)}$. This approach propagates uncertainty in the segmentation without introducing department cluster assignments as additional latent variables, which would substantially increase model complexity and computational cost. Based on $\mathbf{P}$, we obtain a point estimate of the partition using the function \texttt{Mclust} from the \texttt{mclust} package in \texttt{R}. In particular, we apply Gaussian model based clustering to $\mathbf{P}$ and select the number of clusters using the Bayesian information criterion. The resulting maximum a posteriori classification defines the estimated departmental partition \citep{haughton2009review}.

\begin{figure}[!htb]
    \centering
    \includegraphics[width=0.87\textwidth,trim={0cm 0cm 0cm 1.5cm},clip]{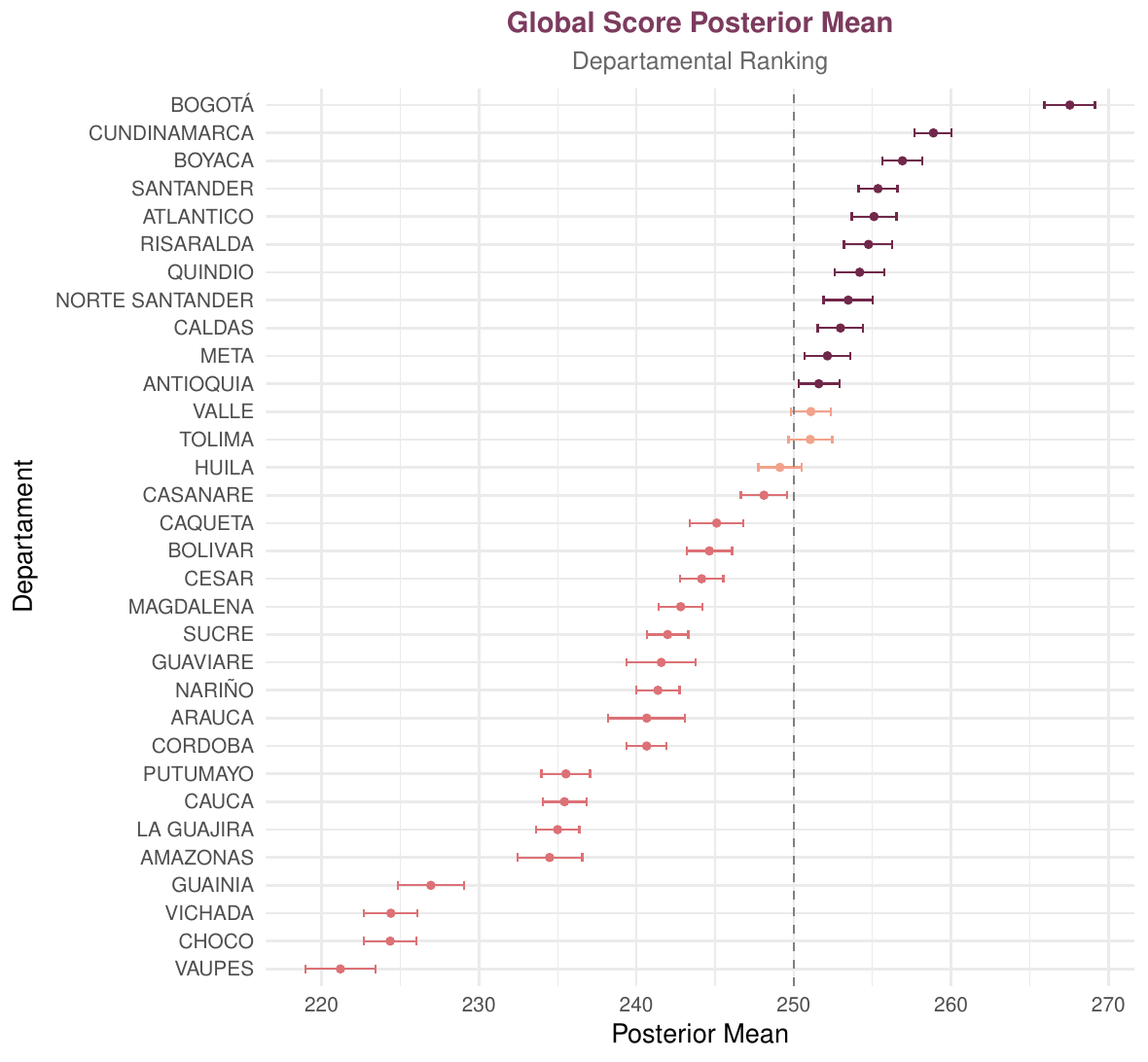}
    \caption{Departmental ranking of the global score posterior mean with 95\% credible intervals. Colors indicate whether the interval lies above 250, includes 250, or lies below 250.}
    \label{fig:global_dep_rank}
\end{figure}

\begin{figure}[!htb]
    \centering
    \subfigure[Departmental coclusering matrix.]
    {\includegraphics[width=0.49\textwidth,trim={0cm 0cm 0cm 2.5cm},clip]{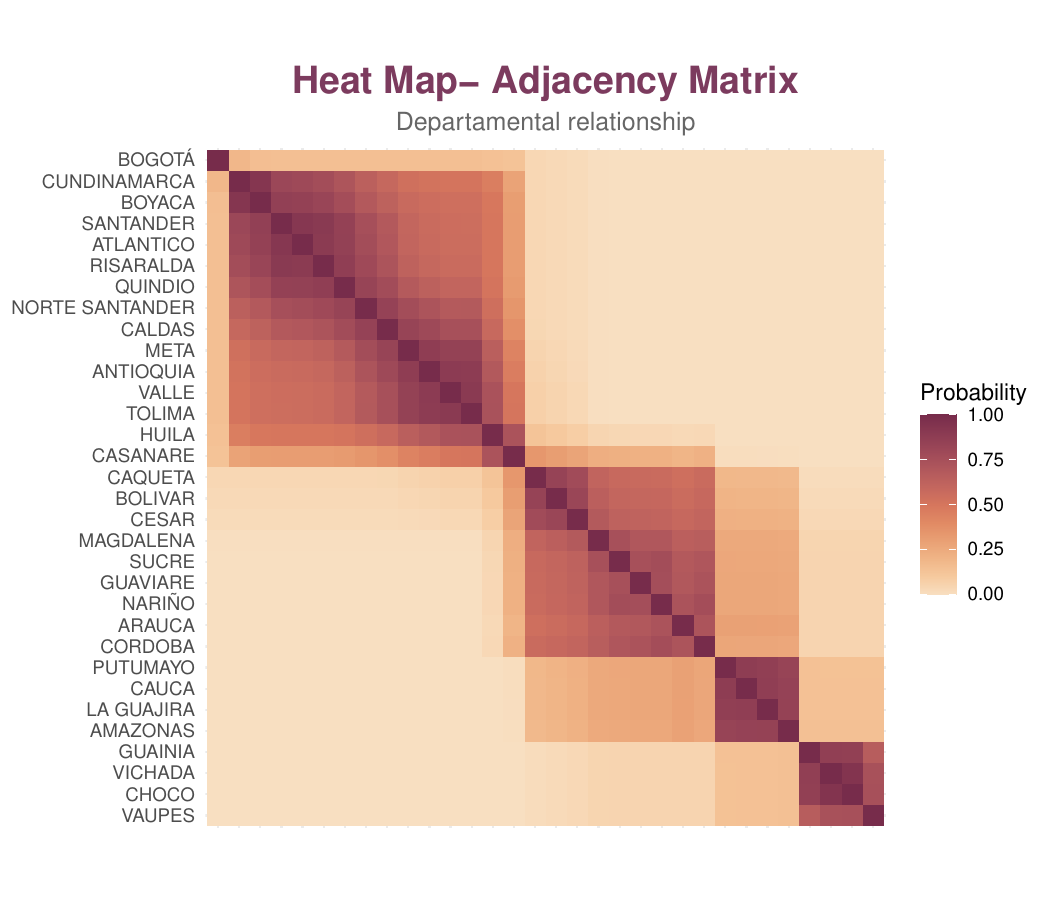}}
    \subfigure[Departmental segmentation.]
    {\includegraphics[width=0.49\textwidth,trim={0cm 0cm 0cm 1.5cm},clip]{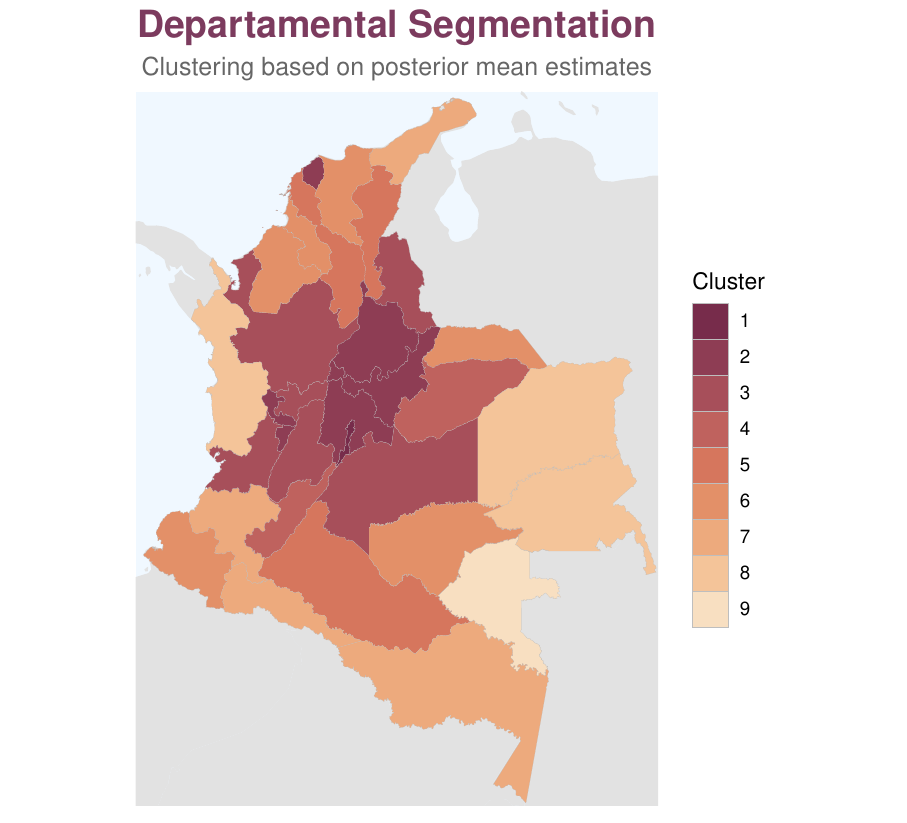}}
    \caption{Departmental clustering summaries based on posterior co-clustering probabilities and the resulting segmentation map.}
    \label{fig:dep_clustering_summary}
\end{figure}

The results reported below are based on departmental posterior means under the Ridge model. Figure \ref{fig:global_dep_rank} shows the posterior mean global score and the corresponding 95\% credible interval for each department. Since the Saber 11 exam was designed to have a theoretical mean of 250, intervals are shown in different colors to indicate whether they are above, include, or fall below 250. These results provide additional evidence consistent with educational centralization. In particular, the highest posterior mean score is observed for Bogotá, whereas the five lowest scoring departments correspond to territories that are widely recognized as historically neglected and affected by severe deprivation, including limited access to basic public services such as electricity, sanitation, and potable water.

\begin{figure}[!b]
    \centering
    \includegraphics[width=0.6\linewidth,trim={0cm 0cm 0cm 1.5cm},clip]{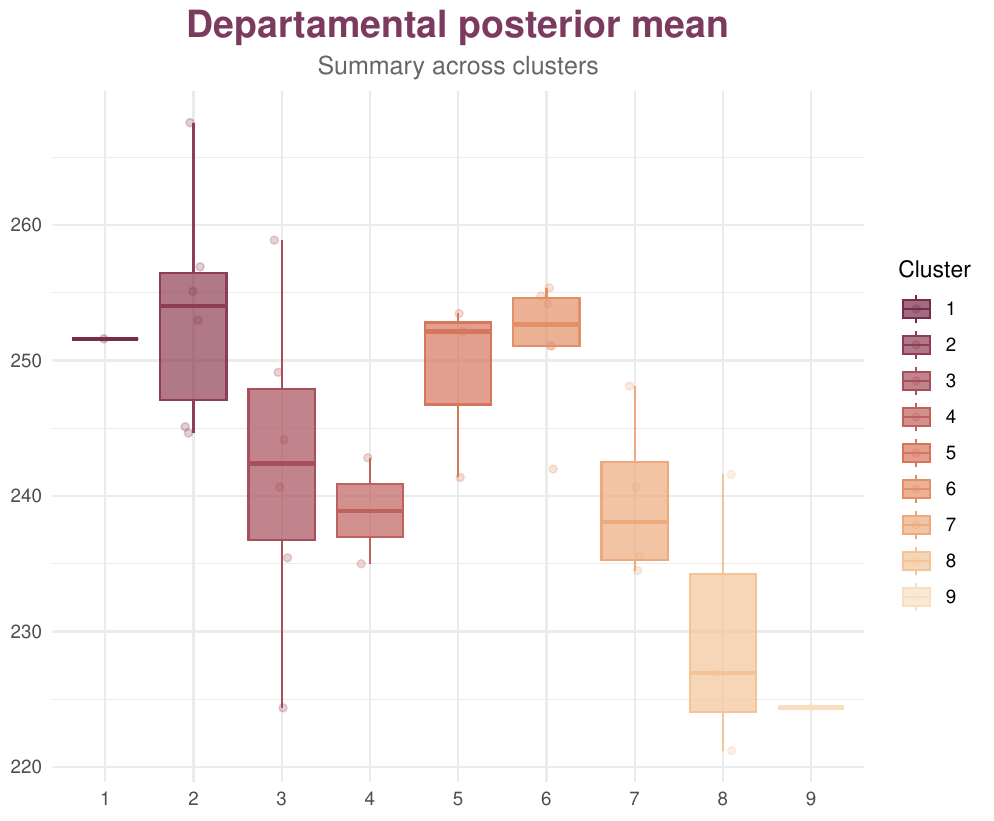}
    \caption{Distribution of departmental posterior mean global scores across clusters.}
    \label{Dep_clus_box_plot}
\end{figure}

Panel (a) of Figure \ref{fig:dep_clustering_summary} shows the heat map of $\mathbf{P}$. Departments are ordered by decreasing posterior global score, which highlights that high performing departments tend to cluster together, whereas low scoring departments form a separate group. Panel (b) shows the corresponding point estimate of the departmental partition on an areal map, with departments colored by cluster. This clustering pattern is consistent with the descriptive statistics from the Saber 11 exam (see Section \ref{sec_eda}). When departments are segmented using posterior means and posterior variability, the highest scores concentrate in the central region of the country. Departments such as Boyacá, Santander, and Cundinamarca exhibit relatively high scores with similar variability, whereas Bogotá, despite having the highest posterior mean, differs in variability and therefore does not cluster with them. As distance from the center increases, posterior global scores tend to decline, which reflects persistent territorial inequalities. Historically marginalized regions, such as Vaupés and Chocó, appear among the lowest scoring departments. In addition, Atlántico and the Coffee Region show higher posterior means, which is consistent with improvements in educational outcomes associated with greater resource availability.

\begin{figure}[!htb]
    \centering
    \includegraphics[width=\linewidth,trim={0cm 0cm 0cm 1.5cm},clip]{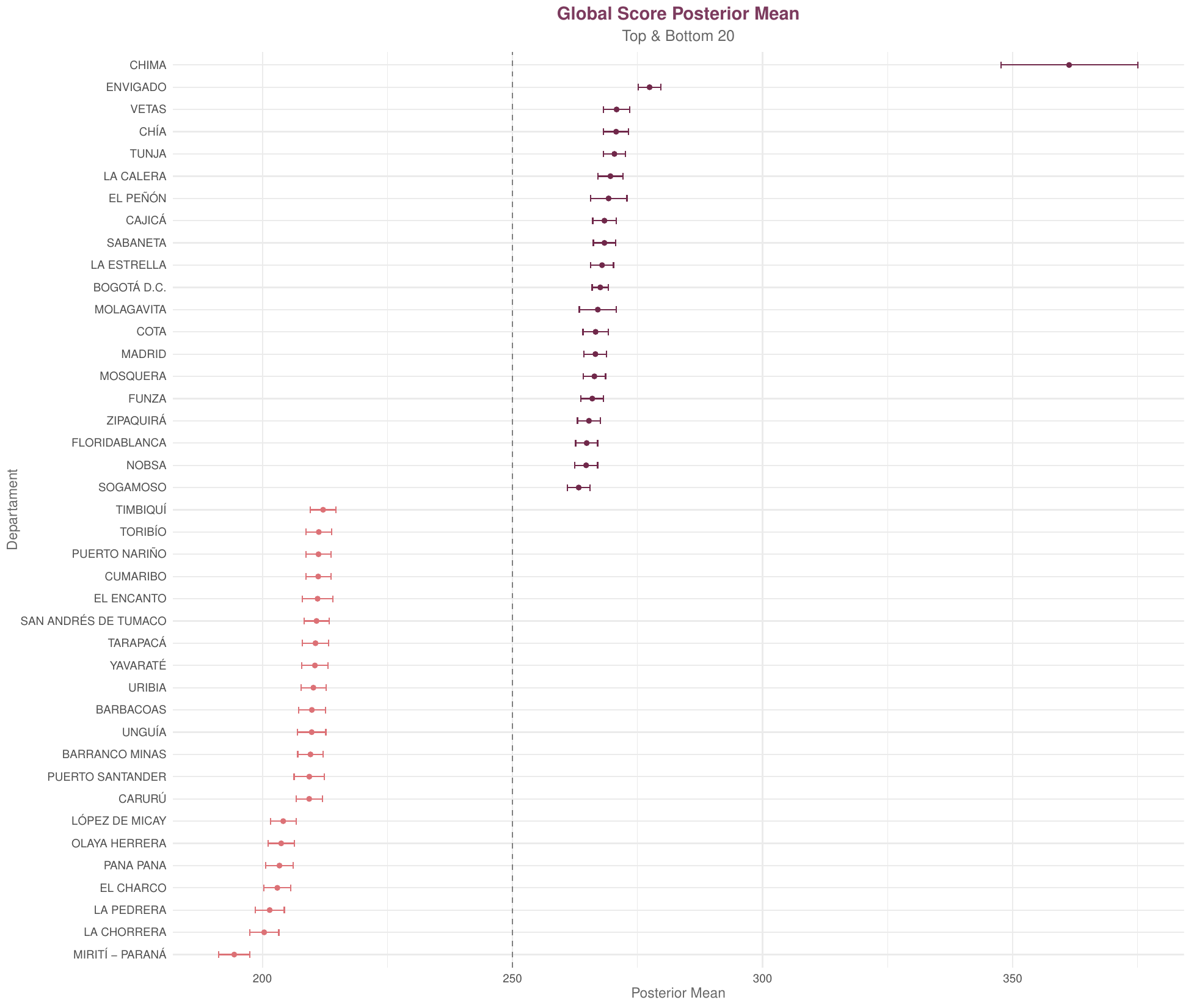}
    \caption{Municipal ranking of the global score posterior mean with 95\% credible intervals, showing the top 20 and bottom 20 municipalities. Colors indicate whether the interval lies above 250, includes 250, or lies below 250.}
    \label{top20_mun_glob_score}
\end{figure}

To further examine the cluster structure, Figure \ref{Dep_clus_box_plot} shows the distribution of departmental posterior means across clusters. Cluster scores are heterogeneous in both location and variability. The first cluster contains only Bogotá, which is expected because its posterior mean is the highest and its posterior variability differs from the remaining departments, so it behaves as an outlier under the clustering criterion. The second cluster exhibits the highest posterior global scores and the largest variability. In contrast, clusters 5 and 6 show comparatively high posterior means with substantially lower variability. The remaining clusters display the lowest global scores, with moderate to high variability.

\begin{figure}[!htb]
    \centering
    \subfigure[Municipal co-clustering matrix.]
    {\includegraphics[width=0.49\textwidth,trim={0cm 0cm 0cm 1.5cm},clip]{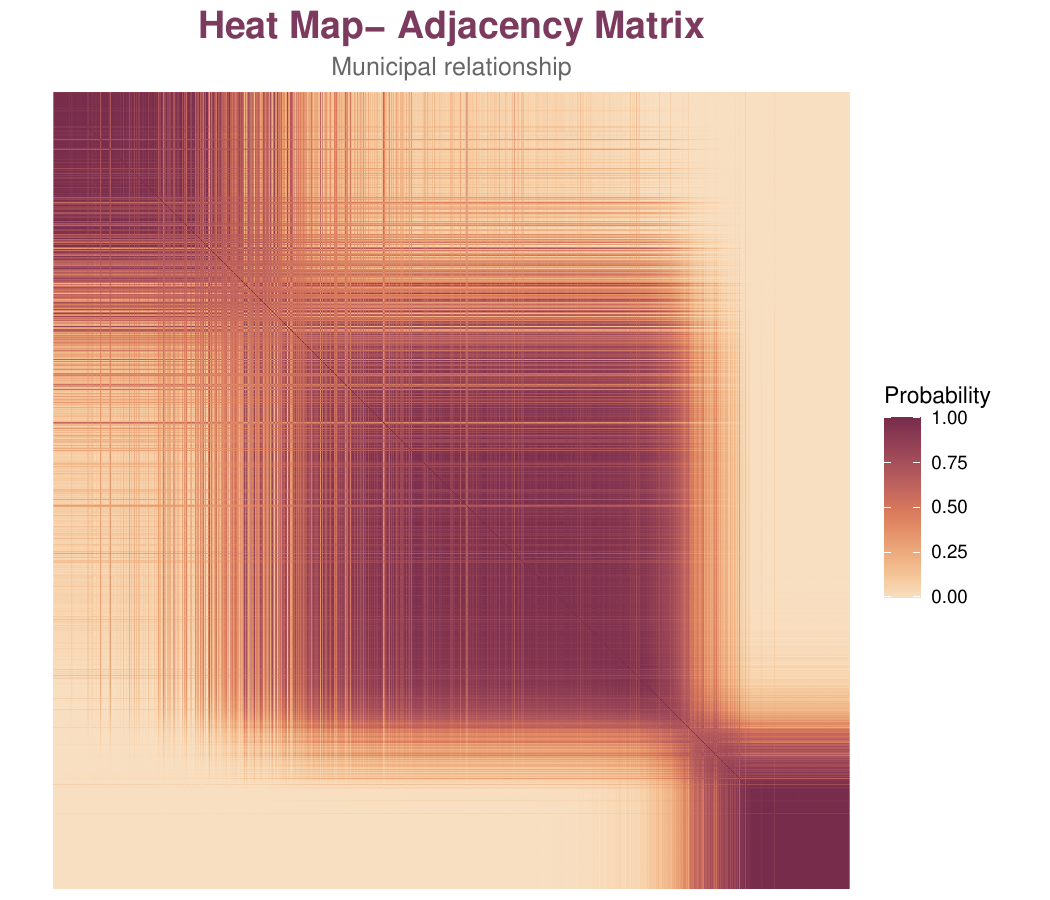}}
    \subfigure[Municipal segmentation.]
    {\includegraphics[width=0.49\textwidth,trim={0cm 0cm 0cm 1.5cm},clip]{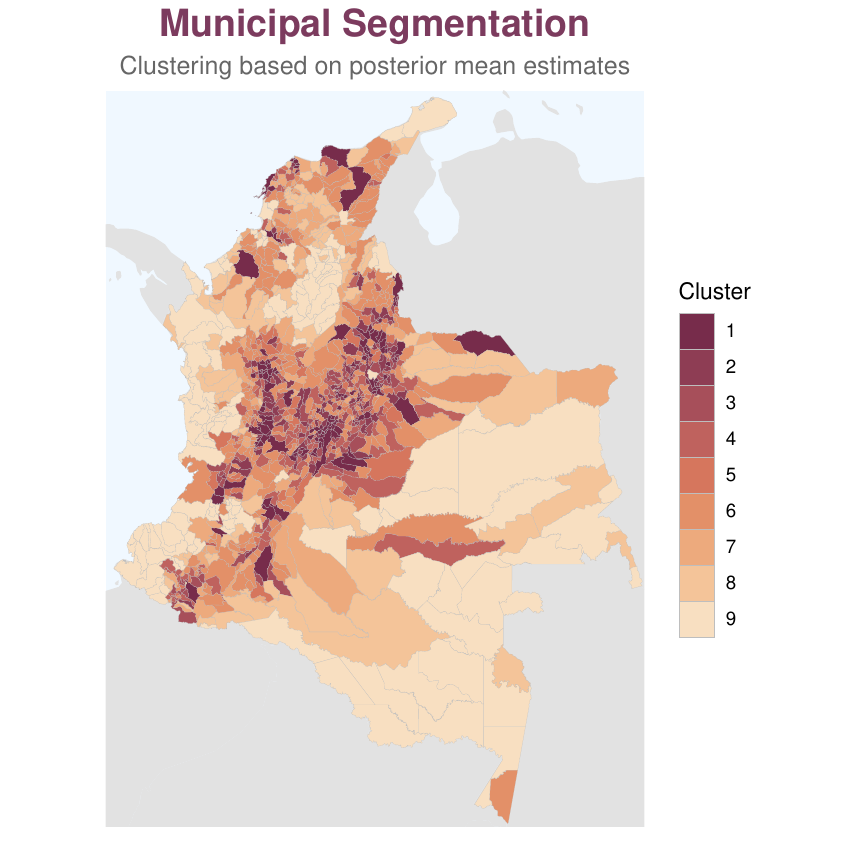}}
    \caption{Municipal clustering summaries based on posterior co-clustering probabilities and the resulting segmentation map.}
    \label{fig:mun_clustering_summary}
\end{figure}

\subsection{Municipal segmentation based on mean global scores}

The municipal segmentation is obtained from municipal posterior means under the Ridge model, using the same co-clustering based strategy adopted at the departmental level provided above. Figure \ref{top20_mun_glob_score} shows the top 20 and bottom 20 municipal posterior means and the corresponding 95\% credible intervals. As in the departmental analysis, municipalities within high scoring departments tend to exhibit higher posterior means. Conversely, municipalities in remote territories and in regions historically characterized by limited state capacity and persistent violence can have posterior means well below 250, in some cases below 200. Beyond location effects, posterior variability reveals additional structure that is not visible from rankings alone.

Panel (a) of Figure \ref{fig:mun_clustering_summary} shows the municipal co-clustering probability matrix, with municipalities ordered by decreasing posterior global score. This ordering highlights that the highest scoring municipalities tend to cluster together, whereas the lowest scoring municipalities form a separate group. Relative to the departmental level, the municipal partition is dominated by three clusters, with a large middle cluster that captures the central range of municipal posterior means. Based on this co-clustering matrix, we obtain a point estimate of the partition using the same procedure adopted at the departmental level. Panel (b) shows the corresponding municipal segmentation on an areal map. The spatial pattern of higher scores around the country’s center is largely preserved, but important exceptions emerge. Some municipalities within departments with low posterior means attain comparatively high scores, which reflects strong territorial inequality and substantial within department heterogeneity.

\begin{figure}[!b]
    \centering
    \includegraphics[width=0.6\linewidth,trim={0cm 0cm 0cm 1.5cm},clip]{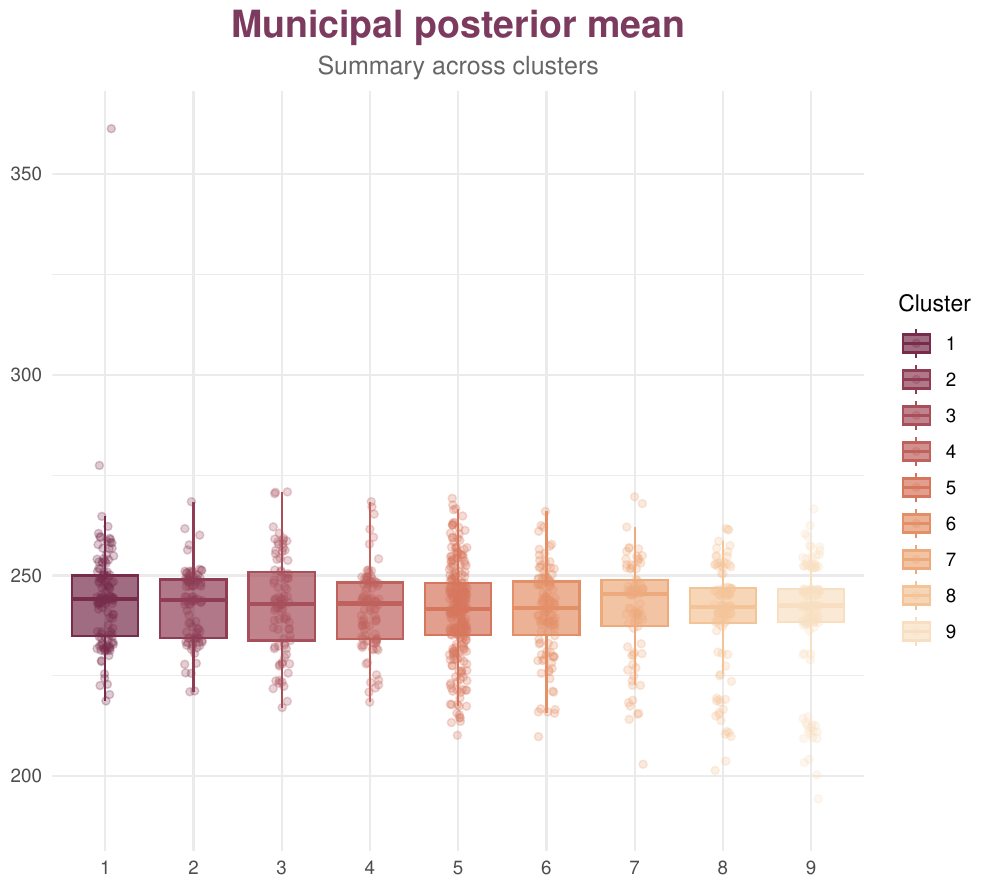}
    \caption{Distribution of municipal posterior mean global scores across clusters.}
    \label{fig:mun_cluster_boxplot}
\end{figure}

To further examine this structure, Figure \ref{fig:mun_cluster_boxplot} summarizes municipal posterior means across clusters. Relative to the departmental segmentation, clusters differ less in location and more in dispersion, which suggests that the municipal partition is driven primarily by differences in heterogeneity and posterior uncertainty rather than by large shifts in average performance. The gradual increase in variance across clusters indicates that municipalities with similar posterior means can still separate into groups with systematically different variability, which is consistent with stronger within cluster heterogeneity in educational conditions, unequal information across municipalities, or both.

\subsection{Model comparison}\label{sec_model_comparison}

To assess model performance, we compare the three model specifications using the Deviance Information Criterion, DIC \citep{spiegelhalter2002bayesian}, and the Watanabe Akaike Information Criterion, WAIC \citep{Watanabe2013Widely}. Both criteria approximate expected out of sample predictive performance while accounting for model complexity through an effective number of parameters, and smaller values indicate better expected predictive ability. The results are reported in Table \ref{tb_DIC_WAIC_real}. Consistent with the simulation study in Section \ref{sec_sim}, the baseline model attains the lowest DIC and WAIC, followed by the Ridge model, whereas the Lasso specification yields the largest values. This pattern is coherent with the higher complexity of the Lasso specification, reflected in a larger estimated effective number of parameters, without a compensating improvement in predictive fit.

\begin{table}[!htb]
\centering
\resizebox{0.85\textwidth}{!}{%
\begin{tabular}{lcccccc}
\toprule
Model & lp & $p_{\text{DIC}}$ & DIC & lppd & $p_{\text{WAIC}}$ & WAIC \\
\midrule
Baseline & 1{,}461.466 & 2{,}466{,}021 & 4{,}929{,}118 & -1{,}231{,}146 &   806.175 & 2{,}463{,}904 \\
Ridge        & 1{,}457.845 & 2{,}467{,}857 & 4{,}932{,}799 & -1{,}231{,}978 &   985.481 & 2{,}465{,}927 \\
Lasso        & 1{,}356.887 & 2{,}501{,}110 & 4{,}999{,}506 & -1{,}247{,}603 & 3{,}280.360 & 2{,}501{,}766 \\
\bottomrule
\end{tabular}%
}
\caption{Information criteria for the fitted Baseline, Ridge, and Lasso models. Here lp is the posterior mean log likelihood, $p_{\text{DIC}}$ and $p_{\text{WAIC}}$ are the effective numbers of parameters, and lppd is the log pointwise posterior predictive density. Smaller values of DIC and WAIC indicate better expected predictive performance.}
\label{tb_DIC_WAIC_real}
\end{table}

Predictive performance is assessed using an external test dataset with 235{,}593 students who took the Saber 11 exam in the second semester of 2022. The test set includes the same municipalities and departments as the training data. For each student in the test set, we generate the posterior predictive distribution of the global score. To obtain a stable point prediction, we implement a resampling scheme that draws 100 values from the posterior predictive distribution and uses their average as the point estimate. This approach reduces Monte Carlo variability and dampens the influence of extreme predictive draws, which is useful when posterior predictive distributions are highly dispersed. As performance metrics, we report the root mean squared error (RMSE), the mean absolute error (MAE), and the coefficient of determination ($R^2$). Table \ref{MAE_RMSE_remuestreo} summarizes out of sample predictive accuracy for the three models. Overall, performance is very similar across specifications. The baseline model attains the smallest RMSE and MAE and the largest $R^{2}$, but the differences are modest. All models achieve $R^{2}$ values between 0.91 and 0.92, which indicates that they explain at least 91\% of the variability in the test set. In addition, MAE remains close to 4 points and RMSE remains between 13 and 16 points, which suggests comparable predictive accuracy across the three shrinkage strategies.

\begin{table}[!htb]
\centering
\begin{tabular}{lccc}
\toprule
Model & RMSE & MAE & $R^{2}$ \\
\midrule
Baseline & 13.31 & 3.68 & 0.92 \\
Ridge    & 15.37 & 4.00 & 0.91 \\
Lasso    & 14.35 & 3.70 & 0.92 \\
\bottomrule
\end{tabular}
\caption{Out of sample predictive performance on a test dataset with 235{,}593 students who took the Saber 11 exam in the second semester of 2022. The test set includes the same municipalities and departments as the training data.}
\label{MAE_RMSE_remuestreo}
\end{table}

\section{Simulation}\label{sec_sim}

We conduct simulation experiments for the three models described in Section \ref{sec_modeling} to evaluate estimation and predictive performance, with emphasis on the role of covariate effects. We generate three synthetic datasets that preserve the original spatial and demographic structure. In particular, we keep fixed the number of students within each municipality and department, as well as the number of municipalities per department. Data generation is based on distinct scenarios for the covariate effects, designed to represent extreme settings and to assess robustness under such conditions. All remaining parameters, including variance components and spatial effects, are fixed at posterior means obtained from the preliminary data analysis in Section \ref{sec_data_analysis} under each model. Table \ref{tab:sim_global_score_summary} reports summary statistics for the simulated global score under each scenario. The distribution shifts upward across scenarios, with increases in quartiles and in the mean, which reflects progressively higher overall performance. Scenario 1 is centered close to 250, whereas Scenarios 2 and 3 exhibit substantially larger central values and wider upper tails.

\begin{table}[!htb]
\centering
\begin{tabular}{lcccccc}
\toprule
Scenario & Min & $Q_{1}$ & $Q_{2}$ & Mean & $Q_{3}$ & Max \\
\midrule
Scenario 1 & 12.94 & 217.65 & 251.3198 & 251.4573 & 285.10 & 497.75 \\
Scenario 2 & 19.45 & 271.76 & 308.7179 & 308.4377 & 345.61 & 544.93 \\
Scenario 3 &  0.24 & 346.28 & 380.5153 & 380.2448 & 414.60 & 602.70 \\
\bottomrule
\end{tabular}
\caption{Summary statistics of the simulated global score by scenario.}
\label{tab:sim_global_score_summary}
\end{table}

We consider three simulation scenarios that vary the magnitude and direction of covariate effects. Scenario 1 assigns moderate coefficients and represents average conditions, so it serves as a baseline configuration. Scenario 2 shrinks most coefficients toward zero to evaluate sensitivity to weak associations. Scenario 3 introduces stronger and contrasting coefficients, including sign reversals for several municipal and departmental covariates, to emulate extreme conditions and induce potential multicollinearity. The parameter values used in each scenario are reported in Tables \ref{tab:sim_student_cov} and \ref{tab:sim_mun_dep_cov}, and they define a deliberately challenging setting for estimation and prediction in Scenario 3.

\begin{table}[!htb]
\centering
\begin{tabular}{lrrr}
\toprule
Student level parameter & Scenario 1 & Scenario 2 & Scenario 3 \\
\midrule
Intercept              & 196.16 & 250.00 & 353.00 \\
Mother's education     &  18.13 &   0.50 &  12.13 \\
Computer access        &  10.36 &   1.00 &  12.00 \\
Internet access        &   8.93 &   1.00 &  12.00 \\
Ethnicity              & -14.34 &  -0.50 &   0.00 \\
Gender                 &  -7.58 &  -0.25 &   0.00 \\
Books (11--25)         &   8.48 &   1.00 &  -9.48 \\
Books (26--100)        &  20.06 &   2.00 &  10.06 \\
Books ($>$100)         &  20.27 &   2.00 &  10.27 \\
Socioeconomic level 1  &  20.36 &  45.00 &   0.00 \\
Socioeconomic level 2  &  18.75 &  45.00 &   0.00 \\
Socioeconomic level 3  &  17.00 &  45.00 &   0.00 \\
Socioeconomic level 4  &  14.45 &  45.00 &   0.00 \\
Socioeconomic level 5  &  11.37 &  45.00 &   0.00 \\
Socioeconomic level 6  &   2.53 &  45.00 &   0.00 \\
Academic calendar A    &  23.81 &   2.00 &  10.80 \\
Academic calendar B    &  19.07 &   1.00 &  19.07 \\
Private school         &  12.82 &   1.00 &  13.82 \\
Work ($<$10 hours)     & -13.42 &  -0.43 & -13.42 \\
Work (11--20 hours)    & -12.85 &  -0.43 & -13.85 \\
Work (21--30 hours)    & -16.16 &  -0.43 & -16.16 \\
Work ($>$30 hours)     & -26.14 &  -0.43 & -26.14 \\
\bottomrule
\end{tabular}
\caption{Simulation scenarios for student level covariate effects.}
\label{tab:sim_student_cov}
\end{table}

For each scenario, we fit the three models using the same MCMC algorithm described in Section \ref{sec_data_analysis}. For each fit, we generate 76{,}500 posterior draws. Before estimation, municipal and departmental covariates are centered to reduce scale effects and improve numerical stability. The first 10\% of draws are discarded as burn in, and we thin the remaining chain every five iterations to reduce autocorrelation, which yields 15{,}000 posterior samples for inference. Convergence is assessed using log likelihood trace plots, effective sample sizes, and Monte Carlo standard errors, and these diagnostics do not indicate lack of convergence. In what follows, we evaluate results across models and scenarios, with emphasis on parameter recovery and predictive performance using the same metrics as in Section \ref{sec_model_comparison}.

\begin{table}[!htb]
\centering
\begin{tabular}{llrrr}
\toprule
Level & Parameter & Scenario 1 & Scenario 2 & Scenario 3 \\
\midrule
Municipal
& Teacher to student ratio        & 2.90  & 0.000 & -5.000   \\
& Victimization risk              & 8.70  & 0.001 & -9.540   \\
& Homicides per 100{,}000         & 0.06  & 0.000 & -0.250   \\
& Students in public schools (\%) & 0.04  & 0.000 & -0.800   \\
& Terrorism index                 & 0.34  & 0.000 & -0.034   \\
& Theft rate                      & 0.00  & 0.000 & -0.000   \\
& Kidnappings                     & 1.12  & 0.100 &  0.800    \\
& Distance to capital city        & 0.05  & 0.001 &  0.005    \\
\midrule
Departmental
& GDP per capita                  & 0.230 & 0.005 & 0.230    \\
& Rural population proportion     & 0.060 & 0.000 & -0.060   \\
& Municipalities at risk (\%)     & 0.004 & 0.000 & -0.040   \\
& Weighted homicides              & 0.012 & 0.000 & -0.120   \\
\bottomrule
\end{tabular}
\caption{Simulation scenarios for municipal and departmental covariate effects.}
\label{tab:sim_mun_dep_cov}
\end{table}

\subsection{Results}

Across the three simulation scenarios, the baseline model without shrinkage shows satisfactory recovery, with 95\% credible intervals covering the true covariate effects in 85.3\%, 70.6\%, and 73.5\% of parameters in Scenarios 1 to 3, respectively, and intervals are generally narrow except for Calendar A and Calendar B. The Ridge specification improves coverage substantially, reaching 94.1\%, 91.1\%, and 79.4\%, while maintaining similarly tight intervals, again with Calendar A and Calendar B as the main exceptions. In contrast, the Lasso specification performs poorly, with coverage of only 32.3\%, 58.8\%, and 50.0\% across scenarios, and it produces wider intervals, which indicates that the strong shrinkage constraint leads to inferior parameter recovery relative to the baseline and Ridge models. For the interested reader, the full set of interval plots underlying these coverage rates is available in the GitHub repository provided in Section \ref{sec_computation}.

Table \ref{tb:DIC_WAIC_sim} summarizes information criteria across the three simulation scenarios. In all scenarios, the Baseline and Ridge specifications attain very similar DIC and WAIC values, with the Baseline model slightly preferred in terms of WAIC and DIC in most cases. In contrast, the Lasso specification performs substantially worse, with markedly smaller lp and lppd, much larger effective complexity measures $p_{\text{DIC}}$ and $p_{\text{WAIC}}$, and therefore much larger DIC and WAIC. This ranking mirrors the empirical application in Table \ref{tb_DIC_WAIC_real}, where the Baseline model achieves the smallest DIC and WAIC, Ridge is close but consistently larger, and Lasso is clearly inferior. Taken together, both the simulated and real data results indicate that shrinkage does not improve predictive fit in this setting, and that Lasso regularization incurs a loss in fit together with a larger effective number of parameters.

\begin{table}[!htb]
\centering
\begin{tabular}{ccccccc}
\toprule
Scenario & lp & $p_{\text{DIC}}$ & DIC & lppd & $p_{\text{WAIC}}$ & WAIC \\
\midrule
\multicolumn{7}{c}{Baseline model} \\
\midrule
1 & 1{,}460.170 & 2{,}468{,}233 & 4{,}933{,}545 & -1{,}232{,}170 &   963.600 & 2{,}466{,}276 \\
2 & 1{,}460.090 & 2{,}467{,}500 & 4{,}932{,}079 & -1{,}231{,}830 &   922.900 & 2{,}465{,}502 \\
3 & 1{,}464.100 & 2{,}467{,}428 & 4{,}931{,}928 & -1{,}231{,}777 &   945.000 & 2{,}465{,}445 \\
\midrule
\multicolumn{7}{c}{Ridge model} \\
\midrule
1 & 1{,}466.556 & 2{,}468{,}383 & 4{,}933{,}833 & -1{,}232{,}167 & 1{,}116.270 & 2{,}466{,}566 \\
2 & 1{,}470.790 & 2{,}467{,}598 & 4{,}932{,}255 & -1{,}231{,}795 & 1{,}066.130 & 2{,}465{,}723 \\
3 & 1{,}470.751 & 2{,}467{,}623 & 4{,}932{,}304 & -1{,}231{,}771 & 1{,}138.300 & 2{,}465{,}819 \\
\midrule
\multicolumn{7}{c}{Lasso model} \\
\midrule
1 & 1{,}366.944 & 2{,}499{,}680 & 4{,}996{,}627 & -1{,}246{,}871 & 3{,}205.150 & 2{,}500{,}152 \\
2 & 1{,}278.309 & 2{,}512{,}537 & 5{,}022{,}518 & -1{,}251{,}350 & 7{,}280.000 & 2{,}517{,}261 \\
3 & 1{,}363.253 & 2{,}492{,}130 & 4{,}981{,}533 & -1{,}242{,}688 & 4{,}026.970 & 2{,}493{,}430 \\
\bottomrule
\end{tabular}
\caption{DIC and WAIC across simulation scenarios for the Baseline, Ridge, and Lasso specifications. Here $lp$ is the posterior mean log likelihood, $pDIC$ and $pWAIC$ are the effective numbers of parameters, and $lppd$ is the log pointwise posterior predictive density. Smaller values of DIC and WAIC indicate better expected predictive performance.}
\label{tb:DIC_WAIC_sim}
\end{table}

Table \ref{tb:MAE_RMSE_sim_all} reports parameter recovery accuracy for covariate effects by hierarchical level. Across scenarios, errors are smallest for municipal and departmental effects and are dominated by student level coefficients, particularly in Scenarios 2 and 3, which are designed to be more challenging. The Baseline and Ridge specifications behave similarly for municipal and departmental parameters, with nearly identical MAE and RMSE in all scenarios, whereas differences are more pronounced at the student level. In Scenario 1, the Baseline model attains the smallest student level errors, while in Scenarios 2 and 3 the Baseline and Ridge models remain comparable, with Ridge yielding slightly smaller RMSE. The Lasso specification performs worse overall, especially for student level parameters in Scenario 1 and for municipal parameters across scenarios, which is consistent with overly aggressive shrinkage and weaker recovery of covariate effects. In the empirical application, Table \ref{MAE_RMSE_remuestreo} also shows small differences in out of sample predictive accuracy across models, with the Baseline model slightly preferred and Ridge close behind, which aligns with the simulation evidence that shrinkage does not yield systematic gains in predictive performance in this setting.

\begin{table}[!htb]
\centering
\begin{tabular}{ccccccc}
\toprule
Level & \multicolumn{2}{c}{Scenario 1} & \multicolumn{2}{c}{Scenario 2} & \multicolumn{2}{c}{Scenario 3} \\
\cmidrule(lr){2-3}\cmidrule(lr){4-5}\cmidrule(lr){6-7}
& MAE & RMSE & MAE & RMSE & MAE & RMSE \\
\midrule
\multicolumn{7}{c}{Baseline model} \\
\midrule
Student       &  0.869 &  1.597 & 20.287 & 23.038 & 14.698 & 34.064 \\
Municipal     &  0.134 &  0.181 &  1.719 &  3.246 &  3.559 &  7.012 \\
Departmental  &  0.075 &  0.117 &  0.005 &  0.005 &  0.113 &  0.135 \\
\midrule
\multicolumn{7}{c}{Ridge model} \\
\midrule
Student       &  1.668 &  3.732 & 19.508 & 22.248 & 14.415 & 32.838 \\
Municipal     &  0.145 &  0.208 &  1.730 &  3.244 &  3.568 &  7.009 \\
Departmental  &  0.075 &  0.117 &  0.005 &  0.005 &  0.113 &  0.135 \\
\midrule
\multicolumn{7}{c}{Lasso model} \\
\midrule
Student       & 15.797 & 18.604 & 13.049 & 23.349 & 12.523 & 24.640 \\
Municipal     &  1.553 &  2.051 &  1.757 &  2.096 &  3.423 &  5.538 \\
Departmental  &  0.077 &  0.119 &  0.002 &  0.002 &  0.112 &  0.134 \\
\bottomrule
\end{tabular}
\caption{Mean absolute error (MAE) and root mean squared error (RMSE) for covariate effects by hierarchical level across simulation scenarios under the Baseline, Ridge, and Lasso specifications.}
\label{tb:MAE_RMSE_sim_all}
\end{table}

\section{Discussion}

This study examines determinants of academic performance in the Saber 11 exam during the second semester of 2022 using three Bayesian hierarchical regression specifications, complemented by simulation scenarios and spatial analyses. Simulation results indicate that Ridge regularization provides the most balanced performance in parameter recovery, predictive accuracy, and sampling efficiency, whereas the Lasso specification yields weaker fit and less stable posterior behavior. From exploratory analysis to posterior rankings and territorial segmentation, a consistent geographic gradient emerges. The highest posterior global scores concentrate in central departments such as Boyacá, Cundinamarca, and Bogotá, with values above the reference mean of 250, whereas lower scores concentrate in peripheral regions such as Chocó, Vaupés, and Vichada. This spatial divide aligns with territorial vulnerability, since several departments with high levels of violence, victimization, and structural deprivation are also located in these peripheral areas.

Posterior estimates show that student level conditions, including mother’s education, access to educational resources, school characteristics, gender, and ethnicity, have the strongest associations with academic outcomes, whereas municipal and departmental covariates exhibit comparatively smaller effects. Even after accounting for these observed factors, spatial random effects retain coherent regional patterns, which supports persistent territorial disparities in educational opportunities across Colombia. Overall, results are consistent with a centralized distribution of educational quality and reveal shortcomings in policy targeting and resource allocation outside major capitals. The persistent disadvantage of rural and historically marginalized territories suggests structural gaps in current strategies and motivates a more territorially oriented educational policy that prioritizes equity through coordinated interventions and sustained public investment.

Beyond the empirical findings, this study develops three new model specifications, together with fully Bayesian computational implementations, tailored to the Colombian context. The proposed framework integrates territorial structure and spatial dependence to uncover latent geographic patterns and to quantify inequalities in access, opportunities, and educational quality. Estimation is carried out with MCMC algorithms that combine Metropolis Hastings updates within a Gibbs sampler. Territorial structure is further summarized by propagating posterior uncertainty into segmentation, applying $k$ means clustering at each iteration and constructing posterior co-clustering probabilities, with segmentation performed at both the departmental and municipal levels to provide a multiscale characterization driven by observed covariates. All results are fully reproducible through the public open source repository referenced in Section \ref{sec_computation}.

Future research offers several directions with practical and methodological impact. On the applied side, the availability of updated and comprehensive territorial information remains limited, which restricts the inclusion of covariates that could better characterize local conditions, particularly those related to armed conflict and violence. Extending the analysis to multiple exam periods and different national administrations would also allow a systematic assessment of how territorial gaps evolve over time. On the methodological side, dynamic hierarchical specifications could be developed to explicitly model temporal evolution in effects and latent structure \citep{west1997bayesian}, and nonparametric Bayes formulations could be explored to better capture unobserved heterogeneity across municipalities and departments \citep{muller2015bayesian}. Given the scale of the data, scalable inference strategies such as variational inference and stochastic variational methods are natural alternatives to full MCMC for rapid experimentation and model comparison \citep{blei2017variational}. Finally, the current segmentation strategy could be replaced or complemented by models that incorporate cluster indicators directly as latent variables, which would enable joint estimation of partitions and regression structure within a unified Bayes framework \citep{Sosa2022BayesianHierarchical}.

\section*{Author Contributions}

\textbf{Laura Pardo}: Data processing, spatial model development, theoretical and computational design of samplers (models with spatial random effects), data analysis and visualization, simulation study, writing and editing.

\textbf{Juan Sosa}: Model formulation and methodology, project supervision and administration, writing and editing.

\textbf{Juan Pablo Torres--Clavijo}: Incorporation of regularization, data extraction and preprocessing, covariates selection based on literature revision, preceding exploratory data analysis and visualization, preceding theoretical and computational design of samplers (models with no spatial random effects).

\textbf{Andrés Felipe Arévalo--Arévalo}: Incorporation of regularization, data extraction and preprocessing, covariates selection based on literature revision, preceding exploratory data analysis and visualization, preceding theoretical and computational design of samplers (models with no spatial random effects).

\section*{Statements and declarations}

The authors declare that they have no known competing financial interests or personal relationships that could have appeared to influence the work reported in this article.

All R and C++ code required to reproduce our results is publicly available at \url{https://github.com/laura-p20/Bayesian-multilevel-model-with-spatial-random-effect}. \\
The repository includes a detailed README with step by step instructions, and the scripts are well documented. All datasets used in the applications and cross validation exercises are also included in the repository.

During the preparation of this work the authors used ChatGPT-5-Thinking in order to improve language and readability. After using this tool, the authors reviewed and edited the content as needed and take full responsibility for the content of the publication.

\bibliography{references.bib}
\bibliographystyle{apalike}

\appendix

\section{Selected proofs}\label{app_proofs}

\subsection{Laplace distribution representation}

The Laplace distribution admits a scale mixture of normals representation with an Exponential mixing distribution. In particular, the marginal prior $\beta \mid \lambda \sim \textsf{Laplace}\left(0, 1/\lambda\right)$ is equivalent to the hierarchical formulation
\[
\beta \mid \tau^2 \sim \textsf{N}\left(0,\tau^2\right),
\quad
\tau^2 \mid \lambda^2 \sim \textsf{Exp}\left(\lambda^2/2\right).
\]
\textbf{Proof:}\\
Let $\beta$ be a random variable with the hierarchical prior
\begin{align*}
\beta \mid \tau^2 &\sim \textsf{N}\left(0,\tau^2\right),
\qquad
\tau^2 \mid \lambda^2 \sim \textsf{Exp}\left(\lambda^2/2\right),
\qquad \lambda>0.
\end{align*}
By the law of total probability, the marginal density of $\beta$ given $\lambda$ is
\begin{align*}
p(\beta \mid \lambda)
&=  \int_{0}^{\infty} p(\beta \mid \tau^2)\,p(\tau^2 \mid \lambda^2)\,\textsf{d}\tau^2 \\
&= \int_{0}^{\infty}
\frac{1}{\sqrt{2\pi\tau^2}}
\exp\left\{-\frac{\beta^2}{2\tau^2}\right\}
\cdot
\frac{\lambda^2}{2}
\exp\left\{-\frac{\lambda^2}{2}\tau^2\right\}
\,\textsf{d}\tau^2 \\
&= \frac{\lambda^2}{2\sqrt{2\pi}}
\int_{0}^{\infty}
(\tau^2)^{\frac{1}{2}-1}
\exp\left\{-\frac{\beta^2}{2\tau^2}-\frac{\lambda^2}{2}\tau^2\right\}
\,\textsf{d}\tau^2.
\end{align*}
The integral is of modified Bessel type. In particular, the identity \citep{gradshteyn2007table}
\begin{align*}
\int_{0}^{\infty} x^{\nu-1}\exp\left\{-\frac{\alpha}{x}-\gamma x\right\}\,\textsf{d}x
=
2\left(\frac{\alpha}{\gamma}\right)^{\nu/2}K_{\nu}\left(2\sqrt{\alpha\gamma}\right),
\quad \alpha>0,\quad \gamma>0,
\end{align*}
applies with $x=\tau^2$, $\nu=\frac{1}{2}$, $\alpha=\frac{\beta^2}{2}$, and $\gamma=\frac{\lambda^2}{2}$. Therefore,
\begin{align*}
p(\beta \mid \lambda)
&= \frac{\lambda^2}{2\sqrt{2\pi}}
\cdot
2\left(\frac{\beta^2}{\lambda^2}\right)^{1/4}
K_{1/2}\left(\lambda|\beta|\right)
=
\frac{\lambda^2}{\sqrt{2\pi}}
\left(\frac{\beta^2}{\lambda^2}\right)^{1/4}
K_{1/2}\left(\lambda|\beta|\right).
\end{align*}
Using $K_{1/2}(z)=\sqrt{\pi/(2z)}\exp(-z)$, we obtain
\begin{align*}
p(\beta \mid \lambda)
=
\frac{\lambda}{2}\,\exp\big(-\lambda|\beta|\big),
\end{align*}
which is the density of a $\textsf{Laplace}\left(0,1/\lambda\right)$ distribution. \qed

\subsection{Ridge regression as a Gaussian prior MAP estimator}

Following the Gaussian sampling model for linear regression $\mathbf{y}\mid \boldsymbol{\beta},\sigma^2 \sim \textsf{N}_{n}\!\left(\mathbf{X}\boldsymbol{\beta},\sigma^2\,\mathbf{I}\right)$, where $\mathbf{X}\in\mathbb{R}^{n\times p}$ and $\boldsymbol{\beta}\in\mathbb{R}^{p}$. For $j=1,\ldots,p$, consider the prior specification
\begin{align*}
\beta_{j}\mid \lambda^{2} \;\overset{\text{iid}}{\sim}\; \textsf{N}\!\left(0,1/\lambda^2\right),
\quad
\lambda^{2}\sim \textsf{G}(a_{\lambda},b_{\lambda}),
\quad
\sigma^{2}\sim \textsf{IG}(a_{\sigma},b_{\sigma}).
\end{align*}
Conditioning on $(\sigma^{2},\lambda^{2})$, the full conditional kernel for $\boldsymbol{\beta}$ is
\begin{align*}
p(\boldsymbol{\beta}\mid -)
&\propto \textsf{N}_{n}\!\left(\mathbf{y}\mid \mathbf{X}\boldsymbol{\beta},\sigma^{2}\mathbf{I}_{n}\right)\,
\prod_{j=1}^{p}\textsf{N}\!\left(\beta_{j}\mid 0,1/\lambda^{2}\right) \\
&\propto \exp\left\{-\frac{1}{2\sigma^{2}}\lVert \mathbf{y}-\mathbf{X}\boldsymbol{\beta}\rVert^{2}-\frac{\lambda^{2}}{2}\boldsymbol{\beta}^{\top}\boldsymbol{\beta}\right\} \\
&= \exp\left\{-\frac{1}{2\sigma^{2}}\left(\lVert \mathbf{y}-\mathbf{X}\boldsymbol{\beta}\rVert^{2}+\lambda^{2}\sigma^{2}\,\boldsymbol{\beta}^{\top}\boldsymbol{\beta}\right)\right\}.
\end{align*}
Therefore, the full conditional distribution is
\begin{align*}
\boldsymbol{\beta}\mid - \sim \textsf{N}_{p}\!\left(\mathbf{m}_{\beta},\mathbf{V}_{\beta}\right),
\quad
\mathbf{V}_{\beta}=\left(\frac{1}{\sigma^{2}}\mathbf{X}^{\top}\mathbf{X}+\lambda^{2}\mathbf{I}_{p}\right)^{-1},
\quad
\mathbf{m}_{\beta}=\mathbf{V}_{\beta}\frac{1}{\sigma^{2}}\,\mathbf{X}^{\top}\mathbf{y}.
\end{align*}
Maximizing $\log p(\boldsymbol{\beta}\mid -)$ is equivalent to minimizing the Ridge objective
\[
\ell_{\text{Ridge}}(\boldsymbol{\beta})=\lVert \mathbf{y}-\mathbf{X}\boldsymbol{\beta}\rVert^{2}+\lambda_{\text{Ridge}}\boldsymbol{\beta}^{\top}\boldsymbol{\beta},
\quad
\lambda_{\text{Ridge}}=\lambda^{2}\sigma^{2},
\]
so the resulting estimator is the maximum a posteriori estimator. \qed

\subsection{Lasso regression as a Gaussian prior MAP estimator}

Following the Gaussian sampling model for linear regression $\mathbf{y}\mid \boldsymbol{\beta},\sigma^2 \sim \textsf{N}_{n}\!\left(\mathbf{X}\boldsymbol{\beta},\sigma^2\,\mathbf{I}\right)$, where $\mathbf{X}\in\mathbb{R}^{n\times p}$ and $\boldsymbol{\beta}\in\mathbb{R}^{p}$. For $j=1,\ldots,p$, consider the prior specification
\begin{align*}
\beta_{j}\mid \tau_j^{2} \;\overset{\text{ind}}{\sim}\; \textsf{N}\!\left(0,\tau_j^{2}\right),
\quad
\tau_j^{2}\mid \lambda^{2}\;\overset{\text{ind}}{\sim}\; \textsf{Exp}\!\left(\lambda^{2}/2\right),
\quad
\lambda^{2}\sim \textsf{G}(a_{\lambda},b_{\lambda}),
\quad
\sigma^{2}\sim \textsf{IG}(a_{\sigma},b_{\sigma}).
\end{align*}
Integrating out $\tau_j^2$ yields the marginal prior $\beta_j\mid \lambda \sim \textsf{Laplace}\!\left(0,1/\lambda\right)$, so $p(\beta_j\mid \lambda)\propto \exp\!\left(-\lambda|\beta_j|\right)$.
Conditioning on $(\sigma^{2},\boldsymbol{\tau}^{2})$, with $\boldsymbol{\tau}^{2}=(\tau_1^2,\ldots,\tau_p^2)^{\top}$, the full conditional kernel for $\boldsymbol{\beta}$ is
\begin{align*}
p(\boldsymbol{\beta}\mid -)
&\propto \textsf{N}_{n}\!\left(\mathbf{y}\mid \mathbf{X}\boldsymbol{\beta},\sigma^{2}\mathbf{I}_{n}\right)\,
\prod_{j=1}^{p}\textsf{N}\!\left(\beta_{j}\mid 0,\tau_j^{2}\right) \\
&\propto \exp\left\{-\frac{1}{2\sigma^{2}}\lVert \mathbf{y}-\mathbf{X}\boldsymbol{\beta}\rVert^{2}
-\frac{1}{2}\sum_{j=1}^{p}\frac{\beta_j^{2}}{\tau_j^{2}}\right\} \\
&= \exp\left\{-\frac{1}{2\sigma^{2}}\left(\lVert \mathbf{y}-\mathbf{X}\boldsymbol{\beta}\rVert^{2}
+\sigma^{2}\boldsymbol{\beta}^{\top}\mathbf{D}_{\tau}^{-1}\boldsymbol{\beta}\right)\right\},
\quad
\mathbf{D}_{\tau}=\textsf{diag}(\tau_1^2,\ldots,\tau_p^2).
\end{align*}
Therefore, the full conditional distribution is
\begin{align*}
\boldsymbol{\beta}\mid - \sim \textsf{N}_{p}\!\left(\mathbf{m}_{\beta},\mathbf{V}_{\beta}\right),
\quad
\mathbf{V}_{\beta}=\left(\mathbf{D}_{\tau}^{-1} +\frac{1}{\sigma^{2}}\mathbf{X}^{\top}\mathbf{X}\right)^{-1},
\quad
\mathbf{m}_{\beta}=\mathbf{V}_{\beta}\frac{1}{\sigma^{2}}\,\mathbf{X}^{\top}\mathbf{y}.
\end{align*}
Moreover, since $\beta_j\mid \lambda \sim \textsf{Laplace}(0,1/\lambda)$, the conditional posterior kernel of $\boldsymbol{\beta}$ given $(\sigma^2,\lambda)$ can be written as
\begin{align*}
p(\boldsymbol{\beta}\mid \mathbf{y},\sigma^2,\lambda)
&\propto \textsf{N}_{n}\!\left(\mathbf{y}\mid \mathbf{X}\boldsymbol{\beta},\sigma^{2}\mathbf{I}_{n}\right)\,
\prod_{j=1}^{p}\textsf{Laplace}\!\left(\beta_j\mid 0,1/\lambda\right) \\
&\propto \exp\left\{-\frac{1}{2\sigma^{2}}\lVert \mathbf{y}-\mathbf{X}\boldsymbol{\beta}\rVert^{2}
-\lambda\sum_{j=1}^{p}|\beta_j|\right\}.
\end{align*}
Maximizing $\log p(\boldsymbol{\beta}\mid \mathbf{y},\sigma^2,\lambda)$ is equivalent to minimizing the Lasso objective
\[
\ell_{\text{Lasso}}(\boldsymbol{\beta})
=\lVert \mathbf{y}-\mathbf{X}\boldsymbol{\beta}\rVert^{2}
+\lambda_{\text{Lasso}}\sum_{j=1}^{p}|\beta_j|,
\quad
\lambda_{\text{Lasso}}=2\sigma^{2}\lambda,
\]
so the resulting estimator is the maximum a posteriori estimator. \qed

\subsection{Identifiability constraint and nonsingularity of the intrinsic CAR precision}

Let $y_i$ denote the response for areal unit $i$. Under the intrinsic CAR specification, and conditional on the remaining areal responses, the full conditional distribution of $y_i$ is
\begin{align*}
y_i \mid \mathbf{y}_{\sim i},\tau^2 \;\overset{\text{ind}}{\sim}\; \textsf{N}\!\left(\frac{1}{d_i}\sum_{j\sim i}y_j,\frac{\tau^2}{d_i}\right),
\end{align*}
where $d_i=\sum_{j=1}^{n} w_{i,j}$ is the number of neighbors of area $i$, and $j\sim i$ denotes that $w_{i,j}=1$ (adjacency).

As recommended by \citet{AG25p}, an identifiability restriction can be imposed on $\mathbf{y}=(y_1,\ldots,y_n)^{\top}$, such as $\sum_{i=1}^{n}y_i=0$.
This restriction is equivalent to projecting $\mathbf{y}$ onto the orthogonal complement of $\mathbf{1}$, since
\[
\mathbf{1}^{\top}\mathbf{y}=\sum_{i=1}^{n}y_i=0.
\]
Under this constraint, the intrinsic precision matrix becomes nonsingular on the constrained subspace. To see this, let $\mathbf{D}=\textsf{diag}(d_1,\ldots,d_n)$ and let $\mathbf{W}=[w_{i,j}]$ be the adjacency matrix. For any nonzero $\mathbf{v}\in\mathbb{R}^{n}$,
\begin{align*}
\mathbf{v}^{\top}(\mathbf{D}-\mathbf{W})\mathbf{v}
&=\sum_{i=1}^{n}d_i v_i^2-\sum_{i=1}^{n}\sum_{j=1}^{n}w_{i,j}v_i v_j
=\frac{1}{2}\sum_{i=1}^{n}\sum_{j=1}^{n}w_{i,j}(v_i-v_j)^2
\geq 0.
\end{align*}
Moreover, $\mathbf{v}^{\top}(\mathbf{D}-\mathbf{W})\mathbf{v}=0$ holds if and only if $v_i=v_j$ whenever $w_{i,j}=1$, which implies $\mathbf{v}\in\textsf{span}(\mathbf{1})$ when the adjacency graph is connected. Since the constraint $\mathbf{1}^{\top}\mathbf{y}=0$ restricts $\mathbf{y}$ to the subspace orthogonal to $\mathbf{1}$, there is no nonzero $\mathbf{v}$ in that subspace for which $\mathbf{v}^{\top}(\mathbf{D}-\mathbf{W})\mathbf{v}=0$. Consequently, $(\mathbf{D}-\mathbf{W})$ is positive definite on the constrained subspace, and therefore it is nonsingular there. \qed

\section{Full conditional distributions}\label{app_fcds}

\subsection{Baseline model}

\begin{itemize}
    \item $\beta\mid-\sim\mathsf{N}\left(\mu,\sigma^2\right),$ with  
    \begin{align*}
&&\mu=\frac{R_{\cdot\cdot}^{(\beta)}+\frac{\mu_\beta}{\sigma^2_\beta}}{\displaystyle\sum_{k=1}^d\sum_{j=1}^{m_k}\frac{n_{jk}}{\kappa^2_{jk}}+\frac{1}{\sigma^2_\beta}}, \quad \sigma^2=\frac{1}{\displaystyle\sum_{k=1}^d\sum_{j=1}^{m_k}\frac{n_{jk}}{\kappa^2_{jk}}+\frac{1}{\sigma^2_\beta}} \quad\text{where}\quad R_{\cdot\cdot}^{(\beta)}=\sum_{k=1}^d\sum_{j=1}^{m_k}\frac{1}{\kappa^2_{jk}}\boldsymbol{r}_{jk}^{(\beta)}\bm{1}_{n_{jk}},
\end{align*}
and $ \boldsymbol{r}_{jk}^{(\beta)}=\boldsymbol{y}_{jk}-\bm{X}_{j,k}\bm{\beta}_\text{E}-\bm{1}_{n_{jk}}\mathbf{z}_{j,k}^{\top}\bm{\beta}_\text{M}-\bm{1}_{n_{jk}}\mathbf{w}_{k}^{\top}\bm{\beta}_\text{D}-\bm{1}_{n_{jk}}\phi_{jk}$.\\
\item $\bm{\beta}_\text{E}\mid-\sim\mathsf{N}_{p_\text{E}}\left(\boldsymbol{\mu},\bm{\Sigma}\right)$, with
\begin{align*}
    &&\bm{\mu}&=\left(\frac{1}{\sigma^2_\text{E}}\textbf{I}+\sum_{k=1}^d\sum_{j=1}^{m_k}\frac{1}{\kappa^2_{jk}}\bm{X}_{jk}^{\top}\bm{X}_{jk}\right)^{-1}\left(\frac{1}{\sigma^2_\text{E}}\bm{\mu}_\text{E} + \sum_{k=1}^d\sum_{j=1}^{m_k}\frac{1}{\kappa^2_{jk}}\bm{X}_{jk}^{\top}\boldsymbol{r}_{jk}^{(\text{E})}\right),\\&&\bm{\Sigma}&=\left(\frac{1}{\sigma^2_\text{E}}\textbf{I}+\sum_{k=1}^d\sum_{j=1}^{m_k}\frac{1}{\kappa^2_{jk}}\bm{X}_{jk}^{\top}\bm{X}_{jk}\right)^{-1},
\end{align*}
where $\boldsymbol{r}_{jk}^{(\text{E})}=\boldsymbol{y}_{jk}-\bm{1}_{n_{jk}}\beta-\bm{1}_{n_{jk}}\mathbf{z}_{jk}^{\top}\bm{\beta}_\text{M}-\bm{1}_{n_{jk}}\mathbf{w}_{k}^{\top}\bm{\beta}_\text{D}-\bm{1}_{n_{jk}}\phi_{jk}$,

\item $\bm{\beta}_\text{M}\mid-\sim\mathsf{N}_{p_\text{M}}\left(\boldsymbol{\mu},\bm{\Sigma}\right),$ with

\begin{align*}
    &&\bm{\mu}&=\left(\frac{1}{\sigma^2_\text{M}}\textbf{I}+\sum_{k=1}^d\sum_{j=1}^{m_k}\frac{n_{jk}}{\kappa^2_{jk}}\textbf{z}_{jk}^{\top}\textbf{z}_{jk}\right)^{-1}\left(\frac{1}{\sigma^2_\text{M}}\bm{\mu}_\text{M} + \sum_{k=1}^d\sum_{j=1}^{m_k}R_{jk}^{(\text{M})}\textbf{z}_{jk}\right),\\&&\bm{\Sigma}&=\left(\frac{1}{\sigma^2_\text{M}}\textbf{I}+\sum_{k=1}^d\sum_{j=1}^{m_k}\frac{n_{jk}}{\kappa^2_{jk}}\textbf{z}_{jk}^{\top}\textbf{z}_{jk}\right)^{-1} \quad\text{where}\quad R_{jk}^{(\text{M})}=\frac{1}{\kappa^2_{jk}}\boldsymbol{r}_{jk}^{(\text{M})\top}\bm{1}_{n_{jk}},
\end{align*}
and $\boldsymbol{r}_{jk}^{(\text{M})}=\boldsymbol{y}_{jk}-\bm{1}_{n_{jk}}\beta-\bm{X}_{jk}\bm{\beta}_\text{E}-\bm{1}_{n_{jk}}\textbf{w}_{k}^{\top}\bm{\beta}_\text{D}-\bm{1}_{n_{jk}}\phi_{jk}.$\\

\item $\bm{\beta}_\text{D}\mid-\sim\mathsf{N}_{p_\text{D}}\left(\boldsymbol{\mu},\bm{\Sigma}\right),$ with
\begin{align*}
    &&\bm{\mu}&=\left(\frac{1}{\sigma^2_\text{D}}\textbf{I}+\sum_{k=1}^d\sum_{j=1}^{m_k}\frac{n_{jk}}{\kappa^2_{jk}}\textbf{w}_{k}^{\top}\textbf{w}_{k}\right)^{-1}\left(\frac{1}{\sigma^2_\text{D}}\bm{\mu}_\text{D} + \sum_{k=1}^d\sum_{j=1}^{m_k}R_{jk}^{(\text{D})}\textbf{w}_{k}\right),\\&&\bm{\Sigma}&=\left(\frac{1}{\sigma^2_\text{D}}\textbf{I}+\sum_{k=1}^d\sum_{j=1}^{m_k}\frac{n_{jk}}{\kappa^2_{jk}}\textbf{w}_{k}^{\top}\textbf{w}_{k}\right)^{-1}\quad\text{where}\quad R_{jk}^{(\text{D})}=\frac{1}{\kappa^2_{jk}}\boldsymbol{r}_{jk}^{(\text{D})\top}\bm{1}_{n_{jk}},
\end{align*}
and $\boldsymbol{r}_{jk}^{(\text{D})}=\boldsymbol{y}_{jk}-\bm{1}_{n_{jk}}\beta-\bm{X}_{jk}\bm{\beta}_\text{E}-\bm{1}_{n_{jk}}\textbf{z}_{jk}^{\top}\bm{\beta}_\text{M}-\bm{1}_{n_{jk}}\phi_{jk}$.\\

\item Let fixed $j$ and $k$, then $\phi_{jk}\mid-\sim\mathsf{N}\left(\mu,\sigma^2\right)$, with
\begin{align*}
    && \mu=\frac{\frac{1}{\tau^2_\phi}\sum_{l\sim j}\phi_{lk}+R^{(\phi)}_{jk}}{\frac{d_{jk}}{\tau^2_\phi}+\frac{n_{jk}}{\kappa^2_{jk}}},\quad\sigma^2=\frac{1}{\frac{d_{jk}}{\tau^2_\phi}+\frac{n_{jk}}{\kappa^2_{jk}}}, \quad\text{where}\quad R^{(\phi)}_{jk}=\frac{1}{\kappa^2_jk}\boldsymbol{r}_{jk}^{(\phi)\top}\bm{1}_{n_{jk}},
\end{align*}
and $\boldsymbol{r}_{jk}^{(\phi)}=\boldsymbol{y}_{jk}-\bm{1}_{n_{jk}}\beta-\bm{X}_{jk}\bm{\beta}_\text{E}-\bm{1}_{n_{jk}}\textbf{z}_{jk}^{\top}\bm{\beta}_\text{M}-\bm{1}_{n_{jk}}\textbf{w}_{k}^{\top}\bm{\beta}_\text{D}.$\\

\item Let fixed $j$ and $k$, then  $\kappa^2_{jk}\mid-\sim\mathsf{IG}\left(\frac{\nu_\kappa+n_{jk}}{2},\frac{\nu_\kappa\kappa^2_k+(\bm{y}_{jk}-\bm{\zeta}_{jk})^{\top}(\bm{y}_{jk}-\bm{\zeta}_{jk})}{2}\right)$.\\

\item Let fixed $k$, then $\kappa^2_{k}\mid-\sim\mathsf{G}\left(\frac{\alpha_\kappa+m_k\nu_\kappa}{2},\frac{\beta_\kappa+\sum_{j=1}^{m_k}\frac{\nu_\kappa}{\kappa^2_{jk}}}{2}\right).$\\

\item $\sigma^2_\beta\mid-\sim\mathsf{IG}\left(\frac{\nu_\beta+1}{2},\frac{\nu_\beta\gamma^2_\beta+(\beta-\mu_\beta)^2}{2}\right)$.\\

\item $\sigma^2_\text{E}\mid-\sim\mathsf{IG}\left(\frac{\nu_\text{E}+p_e}{2},\frac{\nu_\text{E}\gamma^2_\text{E}+(\bm{\beta}_\text{E}-\bm{\mu}_\text{E})^{\top}(\bm{\beta}_\text{E}-\bm{\mu}_\text{E})}{2}\right).$\\

\item  $\sigma^2_\text{M}\mid-\sim\mathsf{IG}\left(\frac{\nu_\text{M}+p_m}{2},\frac{\nu_\text{M}\gamma^2_\text{M}+(\bm{\beta}_\text{M}-\bm{\mu}_\text{M})^{\top}(\bm{\beta}_\text{M}-\bm{\mu}_\text{M})}{2}\right).\\$

\item $\sigma^2_\text{D}\mid-\sim\mathsf{IG}\left(\frac{\nu_\text{D}+p_d}{2},\frac{\nu_\text{D}\gamma^2_\text{D}+(\bm{\beta}_\text{D}-\bm{\mu}_\text{D})^{\top}(\bm{\beta}_\text{D}-\bm{\mu}_\text{D})}{2}\right).$\\

\item $\tau^2_\phi\mid-\sim\mathsf{IG}\left(\frac{\sum_{k=1}^dm_k+\nu_\phi}{2},\frac{\nu_\phi\gamma_\phi+\sum_{k=1}^d\bm{\phi}_k^\top(\bm{D}_k-\bm{\mathcal{W}})\bm{\phi}_k}{2}\right)$.\\

\item $\beta_\kappa\mid-\sim\mathsf{G}\left(\frac{d\alpha_\kappa}{2}+a_{\alpha_\kappa},\frac{\sum_{k=1}^d\kappa^2_k}{2}+b_{\beta_\kappa}\right)$.\\

\item $p(\alpha_\kappa\mid-)\propto\prod_{k=1}^d\frac{(\beta_\kappa)^{\frac{\alpha_\kappa}{2}}}{\Gamma(\frac{\alpha}{2})}(\kappa^2_k)^{(\frac{\alpha_\kappa}{2}-1)}\exp\left\{-\frac{\beta_\kappa}{2}\kappa^2_k\right\}\alpha_\kappa^{(a_{\alpha_\kappa}-1)}\exp\left\{-b_{\alpha_\kappa}\alpha_\kappa\right\}$.

\end{itemize}

\subsection{Ridge model}
For the Ridge model, the full conditional distributions remain the same as baseline model, except for those of the regression coefficients and the parameters involved in their prior specification.
\begin{itemize}
    \item $\bm{\beta}_\text{E}\mid-\sim\mathsf{N}_{p_\text{E}}\left(\boldsymbol{\mu},\bm{\Sigma}\right),$ with
    \begin{align*}
    &&\bm{\mu}&=\left(\lambda^2_\text{E}\textbf{I}+\sum_{k=1}^d\sum_{j=1}^{m_k}\frac{1}{\kappa^2_{jk}}\bm{X}_{jk}^{\top}\bm{X}_{jk}\right)^{-1}\left(\sum_{k=1}^d\sum_{j=1}^{m_k}\frac{1}{\kappa^2_{jk}}\bm{X}_{jk}^{\top}\boldsymbol{r}_{jk}^{(\text{E})}\right),\\
    &&\bm{\Sigma}&=\left(\frac{1}{\sigma^2_\text{E}}\textbf{I}+\sum_{k=1}^d\sum_{j=1}^{m_k}\frac{1}{\kappa^2_{jk}}\bm{X}_{jk}^{\top}\bm{X}_{jk}\right)^{-1},
\end{align*}

\item $\bm{\beta}_\text{M}\mid-\sim\mathsf{N}_{p_\text{M}}\left(\boldsymbol{\mu},\bm{\Sigma}\right),$ with 
\begin{align*}
    &&\bm{\mu}&=\left(\lambda^2_\text{M}\textbf{I}+\sum_{k=1}^d\sum_{j=1}^{m_k}\frac{n_{jk}}{\kappa^2_{jk}}\textbf{z}_{jk}^{\top}\textbf{z}_{jk}\right)^{-1}\left( \sum_{k=1}^d\sum_{j=1}^{m_k}R_{jk}^{(\text{M})}\textbf{z}_{jk}\right),\\
    &&\bm{\Sigma}&=\left(\lambda^2_\text{M}\textbf{I}+\sum_{k=1}^d\sum_{j=1}^{m_k}\frac{n_{jk}}{\kappa^2_{jk}}\textbf{z}_{jk}^{\top}\textbf{z}_{jk}\right)^{-1}.
\end{align*}

\item $\bm{\beta}_\text{D}\mid-\sim\mathsf{N}_{p_\text{D}}\left(\boldsymbol{\mu},\bm{\Sigma}\right),$ with
\begin{align*}
    &&\bm{\mu}&=\left(\lambda^2_\text{D}\textbf{I}+\sum_{k=1}^d\sum_{j=1}^{m_k}\frac{n_{jk}}{\kappa^2_{jk}}\textbf{w}_{k}^{\top}\textbf{w}_{k}\right)^{-1}\left( \sum_{k=1}^d\sum_{j=1}^{m_k}R_{jk}^{(\text{D})}\textbf{w}_{k}\right)\\
    &&\bm{\Sigma}&=\left(\lambda^2_\text{D}\textbf{I}+\sum_{k=1}^d\sum_{j=1}^{m_k}\frac{n_{jk}}{\kappa^2_{jk}}\textbf{w}_{k}^{\top}\textbf{w}_{k}\right)^{-1}.
\end{align*}
\item $\lambda^2_\text{E}\mid-\sim\mathsf{G}\left(\frac{\nu_\text{E}+p_e}{2},\frac{\nu_\text{E}\gamma^2_\text{E}+\bm{\beta}_\text{E}^{\top}\bm{\beta}_\text{E}}{2}\right).$\\

\item $\lambda^2_\text{M}\mid-\sim\mathsf{G}\left(\frac{\nu_\text{M}+p_m}{2},\frac{\nu_\text{M}\gamma^2_\text{M}+\bm{\beta}_\text{M}^{\top}\bm{\beta}_\text{M}}{2}\right).$

\item $\lambda^2_\text{D}\mid-\sim\mathsf{G}\left(\frac{\nu_\text{D}+p_e}{2},\frac{\nu_\text{D}\gamma^2_\text{D}+\bm{\beta}_\text{D}^{\top}\bm{\beta}_\text{D}}{2}\right).$\\

\end{itemize}

\subsection{Lasso model}
For the Lasso model, the full conditional distributions remain the same as baseline model, except for those of the regression coefficients and the parameters involved in their prior specification
\begin{itemize}
    \item $\bm{\beta}_\text{E}\mid-\sim\mathsf{N}_{p_\text{E}}\left(\boldsymbol{\mu},\bm{\Sigma}\right),$ with
    \begin{align*}
    &&\bm{\mu}&=\left(\bm{D}_{\tau^2_\text{E}}+\sum_{k=1}^d\sum_{j=1}^{m_k}\frac{1}{\kappa^2_{jk}}\bm{X}_{jk}^{\top}\bm{X}_{jk}\right)^{-1}\left(\sum_{k=1}^d\sum_{j=1}^{m_k}\frac{1}{\kappa^2_{jk}}\bm{X}_{jk}^{\top}\boldsymbol{r}_{jk}^{(\text{E})}\right),\\
    &&\bm{\Sigma}&=\left(\bm{D}_{\tau^2_\text{E}}+\sum_{k=1}^d\sum_{j=1}^{m_k}\frac{1}{\kappa^2_{jk}}\bm{X}_{jk}^{\top}\bm{X}_{jk}\right)^{-1}\quad\text{where}\quad \bm{D}_{\tau^2_\text{E}}=\text{diag}\left(\frac{1}{\tau^2_{\text{E},\ell}}\right)\quad\ell=1,...,p_{\text{E}}.
\end{align*}

\item $\bm{\beta}_\text{M}\mid-\sim\mathsf{N}_{p_\text{M}}\left(\boldsymbol{\mu},\bm{\Sigma}\right)$, with
\begin{align*}
    &&\bm{\mu}&=\left(\bm{D}_{\tau^2_\text{M}}+\sum_{k=1}^d\sum_{j=1}^{m_k}\frac{n_{jk}}{\kappa^2_{jk}}\textbf{z}_{jk}^{\top}\textbf{z}_{jk}\right)^{-1}\left( \sum_{k=1}^d\sum_{j=1}^{m_k}R_{jk}^{(\text{M})}\textbf{z}_{jk}\right),\\
    &&\bm{\Sigma}&=\left(\bm{D}_{\tau^2_\text{M}}+\sum_{k=1}^d\sum_{j=1}^{m_k}\frac{n_{jk}}{\kappa^2_{jk}}\textbf{z}_{jk}^{\top}\textbf{z}_{jk}\right)^{-1}, \quad\text{where}\quad\bm{D}_{\tau^2_\text{M}}=\text{diag}\left(\frac{1}{\tau^2_{\text{M},r}}\right)\quad r=1,\cdots,p_{\text{M}}.
\end{align*}

    \item $\bm{\beta}_\text{D}\mid-\sim\mathsf{N}_{p_\text{D}}\left(\boldsymbol{\mu},\bm{\Sigma}\right),$ with
    \begin{align*}
    &&\bm{\mu}&=\left(\bm{D}_{\tau^2_{\text{D}}}+\sum_{k=1}^d\sum_{j=1}^{m_k}\frac{n_{jk}}{\kappa^2_{jk}}\textbf{w}_{k}^{\top}\textbf{w}_{k}\right)^{-1}\left( \sum_{k=1}^d\sum_{j=1}^{m_k}R_{jk}^{(\text{D})}\textbf{w}_{k}\right),\\
    &&\bm{\Sigma}&=\left(\bm{D}_{\tau^2_{\text{D}}}+\sum_{k=1}^d\sum_{j=1}^{m_k}\frac{n_{jk}}{\kappa^2_{jk}}\textbf{w}_{k}^{\top}\textbf{w}_{k}\right)^{-1}, \quad\text{where}\quad \bm{D}_{\tau^2_\text{D}}=\text{diag}\left(\frac{1}{\tau^2_{\text{D},t}}\right)\quad t=1,\cdots,p_{\text{D}}.
\end{align*}

\item Let $\ell$ fixed, then $\tau^2_{\text{E},\ell}\simind\mathsf{GIG}(p=\frac{1}{2},a=\lambda^2_\text{E},b=\beta^2_{\text{E},\ell})$.\\
\item Let $r$ fixed, then 
$\tau^2_{\text{M},r}\simind\mathsf{GIG}(p=\frac{1}{2},a=\lambda^2_\text{M},b=\beta^2_{\text{M},r}).$\\
\item Let $t$ fixed, then $\tau^2_{\text{D},t}\simind\mathsf{GIG}(p=\frac{1}{2},a=\lambda^2_\text{D},b=\beta^2_{\text{D},t})$.\\
\item $\lambda^2_\text{E}\mid-\sim\mathsf{G}\left(a_{\lambda_\text{E}}+p_\text{E},\sum_{\ell=1}^{p_\text{E}}\frac{\tau^2_{\text{E},\ell}}{2}+b_{\lambda_\text{E}}\right)$.\\
\item $\lambda^2_\text{M}\mid-\sim\mathsf{G}\left(a_{\lambda_\text{M}}+p_\text{M},\sum_{r=1}^{p_\text{M}}\frac{\tau^2_{\text{M},r}}{2}+b_{\lambda_\text{M}}\right)$.\\
\item $\lambda^2_\text{D}\mid-\sim\mathsf{G}\left(a_{\lambda_\text{D}}+p_\text{D},\sum_{t=1}^{p_\text{D}}\frac{\tau^2_{\text{D},t}}{2}+b_{\lambda_\text{D}}\right)$.

\end{itemize}

\end{document}